\documentclass[english,reprint,nofootinbib,longbibliography,showkeys]{revtex4-1}
\usepackage[utf8]{inputenc}
\usepackage{xcolor}
\usepackage{pdfcolmk}
\usepackage{units}
\usepackage{mathtools}
\usepackage{amsmath}
\usepackage{amssymb}
\usepackage{graphicx}
\PassOptionsToPackage{normalem}{ulem}
\usepackage{ulem}

\makeatletter

\newcommand{\lyxdot}{.}

\providecolor{lyxadded}{rgb}{0,0,1}
\providecolor{lyxdeleted}{rgb}{1,0,0}

\DeclareRobustCommand{\lyxsout}[1]{\ifx\\#1\else\sout{#1}\fi}

\usepackage{dsfont}
\usepackage{graphicx}
\usepackage[english]{babel}
\usepackage{comment}
\usepackage{enumitem}

\usepackage{braket}
\usepackage{amsfonts}
\usepackage{makeidx}
\usepackage{url}
\usepackage[nolist]{acronym}
\usepackage{xcolor}

\usepackage{bbold}

\usepackage[compat=1.1.0]{tikz-feynman}
\newcommand{\propagator}{\begin{tikzpicture} \begin{feynman} \vertex (a) at (0,0); \vertex (b) at (1.2,0) {};          \diagram*{a [particle={\(t\)}] --[plain] b [particle={\(t'\)}] };         \end{feynman}\end{tikzpicture}}
\newcommand{\ghost}{     \begin{tikzpicture} \begin{feynman} \vertex (a) at (0,0); \vertex (b) at (1.2,0) {};          \diagram*{a [particle={\(t\)}] --[charged scalar] b [particle={\(t'\)}] };\end{feynman}\end{tikzpicture}}
\newcommand{\WF}{        \begin{tikzpicture} \begin{feynman} \vertex (a) at (0,0); \vertex [dot] (b) at (1.2,0) {}; \diagram*{a [particle={\(t\)}] --[plain] (b)};         \end{feynman}\end{tikzpicture}}

\newcommand{\diaA}{     \begin{tikzpicture} \begin{feynman} \vertex (a) at (0,0); \vertex [dot] (b) at (0.4,0) {}; \vertex (c) at (0.6,0.35); \vertex (d) at (0.6,-0.35);\diagram*{(a) [particle={\(t_1\)}] --[plain] (b) ; (c) [particle={\(t_2\)}] --[plain] (b) ;  (d) [particle={\(t_3\)}] --[plain] (b)};         \end{feynman}\end{tikzpicture}}

\newcommand{\diaB}{     \begin{tikzpicture} \begin{feynman} \vertex (a) at (0,0); \vertex [dot] (b) at (0.5,0) {}; \vertex (c) at (0.7,0.4); \vertex (d) at (0.7,-0.4); \diagram*{(a) --[plain] (b); (b) --[scalar] (c); (d) --[scalar] (b)};         \end{feynman}\end{tikzpicture}}
\newcommand{\diaC}{     \begin{tikzpicture} \begin{feynman} \vertex (a) at (0,0); \vertex [dot] (b) at (0.5,0) {}; \vertex (c) at (1.0,0.0); \vertex (d) at (0.5,-0.5); \vertex (e) at (0.5,0.5); \diagram*{(a) --[plain] (b); (b) --[plain] (c); (d) --[plain] (b); (e) --[plain] (b)};         \end{feynman}\end{tikzpicture}}

\newcommand{\WFc}{      \begin{tikzpicture} \begin{feynman} \vertex (a) at (0,0); \vertex [dot] (b) at (1.2,0) {}; \diagram*{a --[plain] (b)};         \end{feynman}\end{tikzpicture}}
\newcommand{\propagatorc}{\begin{tikzpicture} \begin{feynman} \vertex (a) at (0,0); \vertex (b) at (1.2,0) {};          \diagram*{a --[plain] b };         \end{feynman}\end{tikzpicture}}

\newcommand{\diaD}{     \begin{tikzpicture} \begin{feynman} \vertex (a) at (0,0); \vertex [dot] (b) at (1.2,0) {}; \vertex [dot] (c) at (1.8,0.7) {}; \vertex [dot] (d) at (1.8,-0.7) {};\diagram*{(a) --[plain] (b) ; (b) --[plain] (c); (b) --[plain] (d)};         \end{feynman}\end{tikzpicture}}
\newcommand{\diaE}{     \begin{tikzpicture} \begin{feynman} \vertex (a) at (0,0); \vertex [dot] (b) at (1.2,0) {}; \vertex (c) at (2.0,0) ;\diagram*{(a) --[plain] (b) ; (b) --[plain, half right] (c); (c) --[plain, half right] (b)};         \end{feynman}\end{tikzpicture}}
\newcommand{\diaF}{     \begin{tikzpicture} \begin{feynman} \vertex (a) at (0,0); \vertex [dot] (b) at (1.2,0) {}; \vertex (c) at (2.0,0) ;\diagram*{(a) --[plain] (b) ; (b) --[scalar, half right] (c); (c) --[scalar, half right] (b)};         \end{feynman}\end{tikzpicture}}
\newcommand{\diaG}{     \begin{tikzpicture} \begin{feynman} \vertex (a) at (0,0); \vertex [dot] (b) at (1.2,0) {}; \vertex [dot] (c) at (1.2,1) {}; \vertex (d) at (2.4,0) ;\diagram*{(a) --[plain] (b) ; (b) --[plain] (c); (b) --[plain] (d)};         \end{feynman}\end{tikzpicture}}

\makeatother

\usepackage{babel}
\begin{document}
\title{Dynamical field inference and supersymmetry}
\author{Margret Westerkamp$^{1,2}$, Igor Ovchinnikov$^{3}$, Philipp Frank$^{1,2}$,
Torsten Enßlin$^{1,2,3}$}
\affiliation{1. Max Planck Institute for Astrophysics, Karl-Schwarzschildstraße
1, 85748 Garching, Germany~\linebreak{}
2. Ludwig-Maximilians-Universität, Geschwister-Scholl Platz 1, 80539
Munich, Germany~\linebreak{}
3. Excellence Cluster Universe, Technische Universität München, Boltzmannstr. 2,
85748 Garching, Germany}
\begin{abstract}
Knowledge on evolving physical fields is of paramount importance in
science, technology, and economics. Dynamical field inference (DFI)
addresses the problem of reconstructing a stochastically driven, dynamically
evolving field from finite data. It relies on information field theory
(IFT), the information theory for fields. Here, the relations of DFI,
IFT, and the recently developed supersymmetric theory of stochastics
(STS) are established in a pedagogical discussion. In IFT, field expectation
values can be calculated from the partition function of the full space-time
inference problem. The partition function of the inference problem
invokes a functional Dirac function to guarantee the dynamics, as
well as a field-dependent functional determinant, to establish proper
normalization, both impeding the necessary evaluation of the path
integral over all field configurations. STS replaces these problematic
expressions via the introduction of fermionic ghost and bosonic Lagrange
fields, respectively. The action of these fields has a supersymmetry,
which means there exists an exchange operation between bosons and
fermions that leaves the system invariant. In contrast to this, measurements
of the dynamical fields do not adhere to this supersymmetry. The supersymmetry
can also be broken spontaneously, in which case the system evolves
chaotically. This affects the predictability of the system and thereby
make DFI more challenging. We investigate the interplay of measurement
constraints with the non-linear chaotic dynamics of a simplified,
illustrative system with the help of Feynman diagrams and show that
the Fermionic corrections are essential to obtain the correct posterior
statistics over system trajectories.
\end{abstract}
\keywords{Information field theory; Field Inference; Supersymmetric Theory of
Stochastics; Stochastic Differential Equations; Chaos Theory}

\maketitle
\begin{acronym}[SUSY] \acro{CG}[CG] {conjugate gradient} \acro{DS}[DS] {dynamical system} \acro{DFI}[DFI] {dynamical field inference} \acro{IFT}[IFT] {information field theory} \acro{KL}[KL] {Kullback-Leibler} \acro{MAP}[MAP] {maximum a posteriori} \acro{ODE}[ODE] {ordinary differential equation} \acro{SDE}[SDE] {Stochastic differential equation} \acro{SEO}[SEO] {stochastic evolution operator} \acro{STS}[STS] {supersymmetric theory of stochastics} \acro{SUSY}[SUSY] {supersymmetry} \acro{QFT}[QFT] {quantum field theory} \acro{QM}[QM] {quantum mechanics} \end{acronym}

\section{Introduction}

\acp{SDE} appear in many disciplines like astrophysics \citep{krulls1994computation},
biology \citep{allen2010introduction}, chemistry \citep{gardiner1985handbook},
and economics \citep{mao2007stochastic,black2019pricing}. In contrast
to traditional differential equations the dynamics of the system,
which follows the \ac{SDE}, are influenced by initial and boundary
conditions but not entirely determined by them. The uncertainty in
the dynamics can be an intrinsic stochastic behavior \citep{uhlenbeck1930theory}
or simply due to imperfections in the model \citep{galenko2010stochastic},
which describes the \ac{DS}.

In addition to the uncertainty introduced by the stochastic process
driving the evolution of the system, any observation of it is noise
afflicted and incomplete. This complicates the inference of the system’s
state further. In previous studies linear \acp{SDE} \citep{https://doi.org/10.1002/andp.202000486},
especially the Langevin SDE \citep{ParisiSourlas}, were already investigated
extensively. Besides this, many numerical methods to solve partial
differential equations were interpreted and the propagation of the
uncertainty for these problems has been studied \citep{SchoberDH2014,doi:10.1098/rspa.2015.0142}.
Here, we consider arbitrary \acp{SDE} and introduce \ac{DFI} as
a Bayesian framework to estimate the state and evolution of a field
following an \ac{SDE} from finite, incomplete, and noise afflicted
data. \ac{DFI} rests on \ac{IFT}, which is information theory
for fields. \ac{IFT} \citep{IFT2,Informationtheoryoffields} was
developed in order to be able to reconstruct an infinite dimensional
signal from some finite dimensional data, as the signal from phy\-si\-cal
reality is usually not limited to the discrete space. Rather a physical
signal is described by a continuous signal field. In contrast the
data taken from a measurement can never be continuous. \ac{IFT} can
then be applied for signal inference in all areas, where limitations
on the exactness of the measurement are given. \ac{DFI} \citep{IFD1,IFD2,IFD3}
utilizes methods from \ac{IFT} for the inference of signals in a
\ac{DS}. The reconstruction of the signal is advanced by the knowledge
on the signal properties, which are specified by the prior covariance
of the signal. Non-linearities in the \ac{SDE} result in a complicated
and signal-dependent structure of the covariance. The central mathematical
object of our investigation will be the partition function of the
inference problem, from which any relevant quantity of interest can
be obtained. The importance of the partition function for the calculation
of dynamical critical properties was also outlined in \citep{Janssen}.
This partition function is represented by a path integral involving
a functional delta function, to enforce the system dynamics, and a
functional determinant, to ensure proper normalization of the involved
probability densities. To handle the delta function and the determinant,
bosonic Lagrange and fermionic ghost fields are respectively introduced.

The approach of \ac{STS} focuses on the theoretical analysis of the
\ac{DS} as a supersymmetric system \citep{Ovchinnikov1,Ovchinnikov2,Ovchinnikov3,Ovchinnikov4}.
One of the central messages of \ac{STS} is the correspondence between
the spontaneous breakdown of this \ac{SUSY} and the emergence of
chaotic dynamics. Here, we argue that the emergence of chaos impacts
the ability to infere dynamical fields. The dynamical growth rates
of the fermionic ghost fields, which are the Lyapuov coefficients
measuring the strength of chaos, impact the uncertainty of any field
inference. Thereby, we illuminate the relevance of central elements
of \ac{STS} for \ac{DFI}.

The paper tries to give a pedagogical introduction into IFT and STS
by presenting the elementary calculation steps in all derivations.
The paper is structured as following: In Sec.\ \ref{sec:Information-Field-Theory-1}
a brief introduction to \ac{IFT} is given, from which's perspective
\ac{DFI} is developed in Sec.\ \ref{sec:Dynamical-Field-Inference}.
Bosonic Lagrange and fermionic ghost fields are introduced in Sec.\ \ref{sec:Dynmical Field Inference with Ghost Fields}.
These permit for a reformulation of the partition function such that
a symmetry between all bosonic and fermionic degrees of freedom becomes
apparent. Sec.\ \ref{sec:SUSY-Breaking-by} investigates the relation
between \ac{SUSY} and \ac{DFI} by showing that system measurements
have no \ac{SUSY} and how spontaneously broken \ac{SUSY} , which
was already investigated in \citep{Witten}, aka chaos impacts field
reconstructions from measurement data. In \ref{subsec:Idealized-Linear-Dynamics}
and \ref{subsec:Idealized-Non-Linear-Dynamics} we analyze the impact
of the chaos on the predictability for linear and non-linear dynamic.
With having connected the \ac{DFI} and \ac{STS} formalisms, and
shown their mutual relevances, we conclude in Sec.\ \ref{sec:Conclusion-and-Outlook}
and give an outlook on future research directions.

\section{Information Field Theory}

\label{sec:Information-Field-Theory-1} In many areas of science,
technology, and economics, the difficult task of interpreting incomplete
and noisy data sets and computing the uncertainty of the results arises
\citep{box2011bayesian,jaynes2003probability}. If the quantity of
interest is a field, for example a spatially extended component of
our Galaxy \citep{leike2019charting,hutschenreuter2020galactic},
or of the atmosphere \citep{stankov2003new,geyer2014high}, which
are mostly continous functions over a physical space, the problem
becomes virtually infinte dimensional, as any point in space-time
carries one or several degrees of freedom. For such problems, which
are called field inference problems, \ac{IFT} was developed. \ac{IFT}
can be considered as a combination of information theory for distributed
quantities and statistical field theory.

\subsection{Notation}

Usually, only certain aspects describing our system $\psi$ are relevant.
These aspects are called the signal, $\varphi$. Phy\-si\-cal degrees
of freedom, which are contained in $\psi$ and not in $\varphi$,
but which still influence the data, are called noise $n$. If $\varphi$
is a physical field $\varphi:\Omega\to\mathbb{R}$, it is a function
that assigns a value to each point in time and $u$-dimensional postition
space. Let us denote a space-time location by $x=(\vec{x},t)\in\Omega=\mathbb{R}^{u}\times\mathbb{R}_{0}^{+}$,
$u\in\mathbb{N}$, where space and time will be handled in the same
manner initially as in \citep{doering1987stochastic,roberts2003step}.
We let the time axis start at $t_{0}=0$ for definiteness.

The field $\varphi=\varphi(x)$ has an infinite number of degrees
of freedom and integrations over the phase space of the field are
represented by path integrals over the integration measure $\mathcal{D}\varphi=\prod_{x\in\Omega}d\varphi_{x}$
\citep{PathIntegrals}, with $\varphi_{x}=\varphi(x)$ being a more
compact notation. In the following these space-time coordinate dependent
fields are denoted as abstract vectors in Hilbert space. The scalar
product between two fields $\varphi(x)$ and $\gamma(x)$ can be written
in short notation as 
\begin{align}
\gamma^{\dagger}\varphi:=\int\text{d}x~\gamma^{*}(x)\varphi(x),\label{eq:ScalarProduct}
\end{align}
where $\gamma^{*}$ is the complex conjugate of $\gamma$, which here
will play no role, as we deal only with real valued fields.

\subsection{Bayesian Updating}

\label{subsubsec:Bayes Updating and the Measurement Process} In order
to get to know a field $\varphi$ one has to measure it. Bayes theorem
states how to update any existing know\-ledge given a finite number
of constraints by measurements that resulted in the data vector $d$.
Apparently, it is not possible to reconstruct the infinite dimensional
field configuration of $\varphi$ perfectly from a finite number of
measurements. This is where the probabilistic description used in
\ac{IFT} comes into play. In probabilistic logic, knowledge states
are described by probability distributions.

After the measurement of data $d$, the knowledge according to Bayes
theorem \citep{Informationtheoryoffields} is given by the posterior
pro\-ba\-bi\-li\-ty distribution 
\begin{align}
\mathcal{P}(\varphi|d)=\frac{\mathcal{P}(d|\varphi)\mathcal{P}(\varphi)}{\mathcal{P}(d)}.\label{eq:Bayes}
\end{align}
This posterior is proportional to the likelihood $\mathcal{P}(d|\varphi)$
of the measured data given the signal field multiplied by the prior
probability distribution $\mathcal{P}(\varphi)$. The normalization
of the posterior is given by the so called evidence 
\begin{align}
\mathcal{P}(d)=\int\mathcal{D}\varphi\,\mathcal{P}(d|\varphi)\,\mathcal{P}(\varphi).
\end{align}
Bayes theorem describes the update of knowledge states. The prior
$\mathcal{P}(\varphi)$ turns into the posterior $\mathcal{P}(\varphi|d)$
given some data $d$. To construct the posterior, we need to have
the prior and the likelihood. The evidence and posterior incorporate
those.

\subsection{Prior knowledge}

The prior probability of $\varphi$, $\mathcal{P}(\varphi)$, specifies
the knowledge on the signal before any measurement was performed.
Formally, the prior on $\varphi$ can be written in terms of the system
prior \citep{IFT2} 
\begin{align}
\mathcal{P}(\varphi)=\int\mathcal{D}\psi~\delta(\varphi-\varphi(\psi))~\mathcal{P}(\psi),
\end{align}
where $\varphi(\psi)$ is the function that specifies the field $\varphi$
given the system state $\psi$. Due to the integration over $\psi$
the underlying system becomes partly invisible in the probability
densities and only the field of interest, the signal field $\varphi$,
remains. Nevertheless, the properties of the original systems will
still be present in the field prior $\mathcal{P}(\varphi)$. For example,
let us consider a situation close to what will be relevant later on.
We consider a system comprised of two interacting fields constituting
the system $\psi=(\varphi,\eta)$, which are related via the invertible
functional $G[\varphi]=\eta$ implying $\mathcal{P}(\eta|\varphi)=\delta(\eta-G[\varphi])$.
Then we have, assuming that there exists a unique solution $\varphi$
to the equation $G[\varphi]=\eta$,
\begin{eqnarray}
\mathcal{P}(\eta|\varphi) & = & \delta(\eta-G[\varphi])\text{ and}\\
\mathcal{P}(\varphi|\eta) & = & \delta(\varphi-G^{-1}[\eta])\nonumber \\
 & = & \frac{\delta(\eta-G[\varphi])}{||\delta G^{-1}[\eta]/\delta\eta||}\nonumber \\
 & = & \left\Vert \frac{\delta G[\varphi]}{\delta\varphi}\right\Vert \ \delta(\eta-G[\varphi]).\label{eq:P(phi|xi)}
\end{eqnarray}
We casted $\mathcal{P}(\varphi|\eta)$ into a form that only requires
to have access to $G$, but not to $G^{-1}$. As $G$ is one-to-one
$\mathcal{P}(\varphi|\eta)=\delta(\varphi-G^{-1}(\eta))$ would be
our preferred quantity to work with. But, in \ac{DFI} of non-linear
systems we rarely have $G^{-1}$ available as an explicit expression
and therefore have to restore to Eq.\ (\ref{eq:P(phi|xi)}). Now,
we assume that we know the prior statistics of $\mathcal{P}(\eta)$
and find the following implications on $\mathcal{P}(\varphi)$,
\begin{align}
\mathcal{P}(\varphi) & =\int\mathcal{D}\eta~\mathcal{P}(\varphi|\eta)\mathcal{P}(\eta)\nonumber \\
 & =\int\mathcal{D}\eta~\left\Vert \frac{\delta G[\varphi]}{\delta\varphi}\right\Vert \ \delta(\eta-G[\varphi])~\mathcal{P}(\eta)\nonumber \\
 & =\left\Vert \frac{\delta G[\varphi]}{\delta\varphi}\right\Vert \ \mathcal{P}(\eta=G[\varphi]).\label{eq:P(phi)}
\end{align}
This shows that the field of interest $\varphi$ inherits the statistics
of the related field $\eta$, however, with a modification by the
functional determinant $||\partial G/\partial\varphi||$ that is sensitive
to non-linearities in the field relation. Here, the probability $\mathcal{P}(\varphi|\eta)$
contains already the two elements that will lead to \ac{SUSY} in
\ac{DFI}, the delta function, which will be represented with bosonic
Lagrange fields and the functional determinant, for which fermionic
fields are introduced. Since both terms contain the functional $G$,
it is plausible that bosons and fermions might be connected via a
symmetry.

\subsection{Likelihood}

Let us now turn to the measurement and its likelihood. The measurement
process of the data can always be written as 
\begin{align}
d=R[\varphi]+n,
\end{align}
if we define the signal response to be $R[\varphi]=\langle d\rangle_{(d|\varphi)}:=\int\mathcal{D}d\,\mathcal{P}(d|\varphi)\,d$
and the noise as $n:=d-R[\varphi]$. In measurement practice, the
response converts a continuous signal into a discrete data set. The
linear noise of the measurement is given by the re\-si\-du\-al vector
in data space between data and the signal response, $n=d-R[\varphi]$.
The statisitcs of the noise, which can be signal dependent, then determines
the likelihood, 
\begin{eqnarray}
\mathcal{P}(d|\varphi) & = & \int\text{\ensuremath{\mathcal{D}n\,\mathcal{P}(d,n|\varphi)\,}}\nonumber \\
 & = & \int\text{\ensuremath{\mathcal{D}n\,\mathcal{P}(d|n,\varphi)\,\mathcal{P}(n|\varphi)\,}}\nonumber \\
 & = & \int\text{\ensuremath{\mathcal{D}n\,\delta(d-R[\varphi]-n)\,\mathcal{P}(n|\varphi)}}\nonumber \\
 & = & \mathcal{P}(n=d-R[\varphi]|\varphi).\label{eq:likelihood}
\end{eqnarray}

Note, however, that we might want to specify initial conditions of
a dynamical field via data as well. Let $\varphi_{0}=\varphi(\cdot,t_{0})$
be the initial field configuration at initial time $t_{0}$. Then,
we specify the initial data to be exactly this initial field configuration,
$d_{0}=\varphi_{0}$, the corresponding response as $R_{0}[\varphi]=\varphi(\cdot,t_{0}),$
and the noise to vanish, $\mathcal{P}(n)=\delta(n)$. Now, the initial
condition is represented via the likelihood $\mathcal{P}(d_{0}|\varphi):=\mathcal{P}(d|\varphi,\,d_{0}\!\!=\!\!\varphi(\cdot,t_{0}))=\delta(\varphi(\cdot,t_{0})-\varphi_{0})$.
This initial data likelihood can be combined with any other data on
the later evolution, $d_{\mathrm{l}}$, via $\mathcal{P}(d|\varphi)=\mathcal{P}(d_{0}|\varphi)\,\mathcal{P}(d_{\mathrm{l}}|\varphi)$,
where $d=(d_{0},d_{\mathrm{l}})$ is the combined data vector.

\subsection{Information}

Bayes theorem Eq.\ (\ref{eq:Bayes}) can be rewritten in terms of
statistical mechanics by defining an information Hamiltonian, or short
the information, which contains all the information needed for inference,
and the partition function, which serves as a normalization factor,
\begin{align}
\mathcal{P}(\varphi|d) & =\frac{e^{-\mathcal{H}(d,\varphi)}}{\mathcal{Z}_{d}},\label{eq:Stat.Mech}\\
\mathcal{H}(d,\varphi) & :=-\ln(\mathcal{P}(d,\varphi)),\label{eq:Hamiltonian}\\
\mathcal{Z}_{d} & :=\int\mathcal{D}\varphi~e^{-\mathcal{H}(d,\varphi)}.\label{eq:GeneralPartitionFunction}
\end{align}
Note, these formal definitions of information Hamiltonian and partition
function hold in the absence of a thermodynamic equilibrium. This
formulation of field inference in terms of a statistical field theory
permits the usage of the well developed apparatus of field theory,
as we briefly show in the following.

\subsection{Partition Function}

\label{subsec:Partition-Function}

There is an infinite number of possible signal field realizations
that meet the constraints given by a finite number of measurements
as encoded in the field posterior $\mathcal{P}(\varphi|d)$. For practical
purposes, for example to have a figure in a publication showing what
is known about a field, one has to extract lower dimensional views
of this very high dimensional posterior function. These can be obtained
by calculating posterior expectation values of the signal field, like
its posterior mean $m=\langle\varphi\rangle_{(\varphi|d)}=\int\mathcal{D}\varphi~\mathcal{P}(\varphi|d)\,\varphi$
or its uncertainty dispersion $D=\langle(\varphi-m)\,(\varphi-m)^{\dagger}\rangle_{(\varphi|d)}$.
Thus, we want to be able to calculate posterior field moments.

Given some data on a signal field $\varphi$ the posterior $n$-point
function is 
\begin{align}
\langle\varphi(x_{1})...\varphi(x_{n})\rangle_{(\varphi|d)}=\int\mathcal{D}\varphi~\varphi(x_{1})~...~\varphi(x_{n})~\mathcal{P}(\varphi|d).
\end{align}
The involved integral can be calculated exactly in case the posterior
$\mathcal{P}(\varphi|d)$ is a Gaussian. Otherwise, the posterior
may be expanded around a Gaussian.

With the help of the moment generating function 
\begin{align}
\mathcal{Z}_{d}[J]=\int\mathcal{D}\varphi~e^{-\mathcal{H}(d,\varphi)+J^{\dagger}\varphi},\label{eq:momentgeneratingfunction}
\end{align}
which incorporates a moment generating source term $J^{\dagger}\varphi=\int\text{d}xJ^{*}(x)\varphi(x)$,
the moments can be calculated via derivation with respect to $J$
as 
\begin{align}
\langle\varphi(x_{1})..\varphi(x_{n})\rangle_{(\varphi|d)}:=\frac{1}{\mathcal{Z}_{d}[J]}\frac{\delta^{n}\mathcal{Z}_{d}[J]}{\delta J^{*}(x_{1})...\delta J^{*}(x_{n})}\bigg\vert_{J=0}.
\end{align}
Likewise the connected correlation functions, also called cumulants,
are defined as 
\begin{align}
\langle\varphi(x_{1})..\varphi(x_{n})\rangle_{(\varphi|d)}^{c}:=\frac{\delta^{n}\log\mathcal{Z}_{d}[J]}{\delta J^{*}(x_{1})...\delta J^{*}(x_{n})}\bigg\vert_{J=0}.\label{eq:Cumulants}
\end{align}
Particularly, the cumulants of first and second order are of importance
as they describe the posterior mean and uncertainty dispersion, $m=\langle\varphi\rangle_{(\varphi|d)}^{\mathrm{c}}=\langle\varphi\rangle_{(\varphi|d)}^{\mathrm{}}$
and $D=\langle\varphi\,\varphi{}^{\dagger}\rangle_{(\varphi|d)}^{\text{c}}=\langle(\varphi-m)\,(\varphi-m)^{\dagger}\rangle_{(\varphi|d)}$,
respectively. Thus, the ultimate goal of any field inference is to
obtain the moment generating partition function $\mathcal{Z}_{d}[J]$
as any desired $n$-point correlation function can be calculated from
it. For this reason, this partition function will be the focus of
our investigations.

\subsection{Free Theory}

\label{subsec:Free Theory} An illustrative example for the signal
reconstruction and the simplest scenario in \ac{IFT} is given by
the free theory. The underlying initial assumptions of the free theory
lead to a theory without non-linear field interactions. In other words,
the information $\mathcal{H}(d,\varphi)$ includes no terms of order
higher than quadratic in the signal field $\varphi$.

The free theory emerges in practice under the following conditions:
\begin{enumerate}
\item[i)] \label{item1:FreeTheory} A Gaussian zero-centered prior, $\mathcal{P}(\varphi)=\mathcal{G}(\varphi,\Phi)$,
with known covariance $\Phi=\langle\varphi\varphi^{\dagger}\rangle_{(\varphi)}$
\item[ii)] \label{item2:FreeTheory} A linear measurement, $d=R~\varphi+n$,
with known linear response $R$ and additive noise
\begin{enumerate}
\item[iii)] \label{item3:FreeTheory }A signal independent Gaussian noise, $\mathcal{P}(n|\varphi)=\mathcal{G}(n,N)$,
with known covariance $N=\langle nn^{\dagger}\rangle_{(n)}$
\end{enumerate}
\end{enumerate}
The information $\mathcal{H}(d,\varphi)$ is then calculated via the
data likelihood and the signal prior, 
\begin{align}
\mathcal{H}(d,\varphi)=-\log(\mathcal{P}(d|\varphi))-\log(\mathcal{P}(\varphi)).
\end{align}
With the assumptions of the free theory and Eq.\ (\ref{eq:likelihood})
the likelihood is 
\begin{align}
\mathcal{P}(d|\varphi) & =\mathcal{G}(R~\varphi-d,N).
\end{align}
Thus, the information for the free theory is given by 
\begin{align}
\mathcal{H}(d,\varphi) & =-\log(\mathcal{G}(R\varphi-d,N)~\mathcal{G}(\varphi,\Phi))\\
 & =\frac{1}{2}\varphi^{\dagger}(R^{\dagger}N^{-1}R+\Phi^{-1})\varphi-d^{\dagger}N^{-1}R\varphi\nonumber \\
 & ~~~~+\frac{1}{2}\ln(|2\pi N|)+\frac{1}{2}\ln(|2\pi\Phi|)+\frac{1}{2}d^{\dagger}N^{-1}d\\
 & =\frac{1}{2}\varphi^{\dagger}D^{-1}\varphi-j^{\dagger}\varphi+\mathcal{H}_{0}.\label{eq:FreeTheoryHammiltonian}
\end{align}
Here, the so called information source $j$, the information propagator
$D$, and $\mathcal{H}_{0}$ were introduced. The latter contains
all the terms of the information that are constant in $\varphi$.
The others are, 
\begin{align}
D & =\left(\Phi^{-1}+R^{\dagger}N^{-1}R\right)^{-1},\label{eq:Dprop}\\
 & =\Phi-\Phi\:R^{\dagger}\left(R\,\Phi\:R^{\dagger}+N\right)^{-1}R\,\Phi\label{eq:Dprop-data-space}\\
j & =R^{\dagger}N^{-1}d.
\end{align}
The second form of the information propagator $D$ can be verified
via explicit calculation,

\begin{eqnarray}
 &  & D\,D^{-1}\nonumber \\
 & = & \left[\Phi-\Phi\:R^{\dagger}\left(R\,\Phi\:R^{\dagger}+N\right)^{-1}R\,\Phi\right]\,\left[\Phi^{-1}+R^{\dagger}N^{-1}R\right]\nonumber \\
 & = & \left[\mathbb{1}-\Phi\:R^{\dagger}\left(R\,\Phi\:R^{\dagger}+N\right)^{-1}R\right]\,\left[\mathbb{1}+\Phi\,R^{\dagger}N^{-1}R\right]\nonumber \\
 & = & \mathbb{1}+\Phi\,R^{\dagger}N^{-1}R-\Phi\:R^{\dagger}\left(R\,\Phi\:R^{\dagger}+N\right)^{-1}R\nonumber \\
 &  & -\Phi\:R^{\dagger}\left(R\,\Phi\:R^{\dagger}+N\right)^{-1}R\,\Phi\,R^{\dagger}N^{-1}R\nonumber \\
 & = & \mathbb{1}+\Phi\,R^{\dagger}N^{-1}R-\Phi\:R^{\dagger}\left(R\,\Phi\:R^{\dagger}+N\right)^{-1}R\nonumber \\
 &  & -\Phi\:R^{\dagger}\left(R\,\Phi\:R^{\dagger}+N\right)^{-1}\left(R\,\Phi\,R^{\dagger}+N-N\right)N^{-1}R\nonumber \\
 & = & \mathbb{1}+\Phi\,R^{\dagger}N^{-1}R-\Phi\:R^{\dagger}\left(R\,\Phi\:R^{\dagger}+N\right)^{-1}R\nonumber \\
 &  & -\Phi\:R^{\dagger}N^{-1}R+\Phi\:R^{\dagger}\left(R\,\Phi\:R^{\dagger}+N\right)^{-1}R\nonumber \\
 & = & \mathbb{1}
\end{eqnarray}
and also holds in the limit $N\rightarrow0$ of a noise-less measurement.

The information can be expressed in terms of the field 
\begin{equation}
m=Dj\label{eq:WFm}
\end{equation}
by completing the square in Eq.~\eqref{eq:FreeTheoryHammiltonian},
which is also known as the generalized Wiener filter solution \citep{wiener1930generalized}.
Also this can be written in a form that permits a noiseless measurement
limit,
\begin{eqnarray}
m & = & \left(\Phi^{-1}+R^{\dagger}N^{-1}R\right)^{-1}R^{\dagger}N^{-1}d\nonumber \\
 & = & R^{\dagger}\Phi\,\left(R\,\Phi\:R^{\dagger}+N\right)d,\label{eq:m-data-space}
\end{eqnarray}
which can be verified with a very analogous calculation.

Only terms, which depend on the signal field $\varphi$ need to be
considered and therefore the symbol ``$\widehat{=}$'' $~$is introduced,
to mark the equality up to an additive constant. We therefore have
\begin{align}
\mathcal{H}(d,\varphi)~\widehat{=}~\frac{1}{2}(\varphi-m)^{\dagger}D^{-1}(\varphi-m).
\end{align}
Knowing the information, the moment generating function of the free
theory, $Z_{\mathcal{G}}[J]$, is constructed in the next step on
the way of calculating the best fit reconstruction of the signal by
means of expectation values. 
\begin{align}
\mathcal{Z}_{\mathcal{G}}[J] & =\int\mathcal{D}\varphi\,e^{-\mathcal{H}(d,\varphi)+J^{\dagger}\varphi}\label{eq:momentGeneratingFunction}\\
 & =\sqrt{|2\pi D|}e^{\frac{1}{2}(j+J)^{\dagger}D(j+J)-\mathcal{H}_{0}}
\end{align}
All higher order ($n>2$) cumulants vanish and the non-vanishing cumulants
are, 
\begin{align}
m(x) & =\langle\varphi(x)\rangle_{(\varphi|d)}^{c}=\frac{\delta\log\mathcal{Z}_{\mathcal{G}}[J]}{\delta J^{*}(x)}\bigg\vert_{J=0},\\
D(x,y) & =\langle\varphi(x)\varphi^{*}(y)\rangle_{(\varphi|d)}^{c}=\frac{\delta^{2}\log\mathcal{Z}_{\mathcal{G}}[J]}{\delta J^{*}(x)\delta J(y)}\bigg\vert_{J=0}.
\end{align}
As higher order cumulants vanish the posterior distribution can be
written as a Gaussian with mean $m$ and uncertainty covariance $D$,
\begin{align}
\mathcal{P}(\varphi|d)=\mathcal{G}(\varphi-m,D).\label{eq:WFposterior}
\end{align}
Hence, computations in free theory are simple, as the Gaussian posterior
can be treated analytically. The usage of the same symbol $D$ for
the information propagator, the inverse of the kernel of the quadratic
term in the information, and the posterior uncertainty dispersion
is justified, as they coincide in the free theory, but only there.

In other cases, when the signal or noise are non-Gaussian, the response
non-linear or the noise is signal dependent, the theory becomes interacting
in the sense that $\mathcal{H}(d,\varphi)$ contains terms that are
of higher than quadratic order. Thus, the information of this non-free,
interacting theory incorporates not only the propagator and source
terms of the free theory but also interaction terms between more than
two signal field values. We will encounter such situations for a field
with non-linear dynamics.

\section{Dynamical Field Inference}

\label{sec:Dynamical-Field-Inference}

\subsection{Field prior}

In the previous section, we saw how to infer a signal field from measurement
data $d$ with some measurement noise $n$ particularly in the case
of a free theory. Now, we consider a \ac{DS}, for which the time
evolution of the signal field is described by an \ac{SDE} 
\begin{align}
\partial_{t}\varphi(x)=F[\varphi](x)+\xi(x).\label{eq:TimeDep.SDE}
\end{align}
We want to see, how this knowledge can be incorporated into a prior
for the field for DFI. The first part of the \ac{SDE} in Eq.~\eqref{eq:TimeDep.SDE},
$\partial_{t}\varphi(x)=F[\varphi](x)$, describes the deterministic
dynamics of the field. The excitation field $\xi$ turns the deterministic
evolution into an \ac{SDE} and mirrors the influence of external
factors on the dynamics. \ac{DFI} aims to infer a signal in such
a \ac{DS} using the tools from \ac{IFT}. Thus, in \ac{DFI}
next to the observational $n$, which results from the measurement
contaminated by nuisance influences, the excitation field $\xi$ of
the \ac{SDE} has to be considered during inference.

Care has to be taken as the domains of the fields $\varphi$ and $\xi$
differ. While $\varphi(x)$ is defined far all $x\in\Omega=\mathbb{R}^{u}\times\mathbb{R}_{0}^{+}$,
the fields $\partial_{t}\varphi$ and $\xi$ live only over $\Omega'=\mathbb{R}^{u}\times\mathbb{R}^{+}$,
from which the intial time slice at $t_{0}=0$ is removed. Eq.\ (\ref{eq:TimeDep.SDE})
therefore makes only statements about fields on $\Omega'$, although
it also depends on the intial conditions $\varphi_{0}=\varphi(\cdot,t_{0}).$
As such need to be specified, a inital condition prior $\mathcal{P}(\varphi_{0})$
is required. We further introduce the notation $\varphi'=\varphi(\cdot,t\neq t_{0})$
for all field degrees of freedom except the ones fixed by the inital
condition, $\varphi_{0}$, so that we have $\varphi=(\varphi_{0},\varphi')$

The \ac{SDE} in Eq.~\eqref{eq:TimeDep.SDE} can be condensed and
generalized by an operator $G[\varphi]$, $G:C^{n,1}(\Omega)\rightarrow C(\Omega')$,
which contains all the time and space derivatives of the SDE up to
order $n$ in space. 
\begin{align}
G[\varphi](x) & =\xi(x)~\text{with}\\
G[\varphi](x) & \coloneqq\partial_{t}\varphi(x)-F[\varphi(\cdot,t)](x)
\end{align}
Within the framework of this study, we will assume that the excitation
of the \ac{SDE} has a prior Gaussian statistics, 
\begin{align}
\mathcal{P}(\xi) & =\mathcal{G}(\xi,\Xi),\label{eq:Excitation}
\end{align}
with known covariance $\Xi$. For a general $G$, $\xi$ in its present
form does not fully specify $\varphi$, for this additionally some
initial conditions $\varphi_{0}$ at time $t_{0}$ have to be specified.
We fix this by augmenting $\xi$ with $\varphi_{0}=\varphi(\cdot,t_{0})$
by setting $\eta=(\varphi_{0},\xi)^{\dagger}$ with
\begin{equation}
\mathcal{P}(\eta)=\mathcal{P}(\varphi_{0})\ \mathcal{G}(\xi,\Xi),
\end{equation}
and by extending $G$ to 
\begin{eqnarray}
G'[\varphi] & = & (\varphi_{0},G[\varphi])
\end{eqnarray}
with $G':C^{n,1}(\Omega)\rightarrow C(\Omega)$ such that $G'[\varphi]=\eta$
and $G'^{-1}[\eta]=\varphi$ hold and are both uniquely defined.

Then, the prior probability for the signal field is accoring to Eq.\ \eqref{eq:P(phi|xi)},
\begin{align}
\mathcal{P}(\varphi) & =\mathcal{P}(\eta=G'[\varphi])\ \left\Vert \frac{\delta G'[\varphi]}{\delta\varphi}\right\Vert \nonumber \\
 & \mathcal{=G}(G[\varphi],\Xi)\,\mathcal{P}(\varphi_{0})\ \left\Vert \frac{\delta G'[\varphi]}{\delta\varphi}\right\Vert ,
\end{align}
and the functional determinant becomes 
\begin{eqnarray}
\left\Vert \frac{\delta G'[\varphi]}{\delta\varphi}\right\Vert  & = & \left\Vert \begin{pmatrix}\frac{\delta\varphi_{0}}{\delta\varphi_{0}} & \frac{\delta G[\varphi]}{\delta\varphi_{0}}\\
\frac{\delta\varphi_{0}}{\delta\varphi'} & \frac{\delta G[\varphi]}{\delta\varphi'}
\end{pmatrix}\right\Vert =\left\Vert \begin{pmatrix}\mathbb{1} & \frac{\delta G[\varphi]}{\delta\varphi_{0}}\\
\mathbb{0} & \frac{\delta G[\varphi]}{\delta\varphi'}
\end{pmatrix}\right\Vert \nonumber \\
 & = & \left\Vert \frac{\delta G[\varphi]}{\delta\varphi'}\right\Vert ,
\end{eqnarray}
where we note that $\delta G/\delta\varphi':C^{n,1}(\Omega)\times C(\Omega')\rightarrow C(\Omega')$
and therefore, after evaluation of this for a specific field configuration
$\varphi$, $\delta G[\varphi]/\delta\varphi':C(\Omega')\rightarrow C(\Omega')$
is a linear operator, which actually is an isomorphism. Thus, we get
finally
\begin{align}
\mathcal{P}(\varphi) & \mathcal{=G}(G[\varphi],\Xi)\,\mathcal{P}(\varphi_{0})\ \left\Vert \frac{\delta G[\varphi]}{\delta\varphi'}\right\Vert .
\end{align}
If we want to have the initial conditions unconstrained, we could
set $\mathcal{P}(\varphi_{0})=\text{const}$. This is possible, as
we could specify initial or later time conditions via additional data
on the field, as explained before.

\subsection{Partition Function}

\ac{DFI} builds on $\mathcal{P}(d,\varphi)=\mathcal{P}(d|\varphi)\,\mathcal{P}(\varphi)$,
the joint probability of data and field, to obtain field expectation
values by investigating the moment generating partition function
\begin{eqnarray}
\mathcal{Z}_{d}[J] & = & \int\mathcal{D}\varphi\,\mathcal{P}(d,\varphi)\,e^{J^{\dagger}\varphi}\nonumber \\
 & = & \int\mathcal{D}\varphi\,\mathcal{P}\left(d|\varphi\right)\,\mathcal{P}(\varphi)\,e^{J^{\dagger}\varphi}\nonumber \\
 & = & \int\mathcal{D}\varphi\,\frac{e^{-\frac{1}{2}\left(d-R\varphi\right)^{\dagger}N^{-1}\left(d-R\varphi\right)+J^{\dagger}\varphi}}{\sqrt{\vert2\pi N\vert}}\ \mathcal{P}(\varphi)\nonumber \\
 & = & \int\mathcal{D}\varphi\,\frac{e^{-\frac{1}{2}\varphi^{\dagger}R^{\dagger}N^{-1}R\varphi+(J+j)^{\dagger}\varphi-\frac{1}{2}d^{\dagger}N^{-1}d}}{\sqrt{\vert2\pi N\vert}}\ \mathcal{P}(\varphi)\nonumber \\
\text{with }j & = & R^{\dagger}N^{-1}d.
\end{eqnarray}
Here, we used that the measurement noise exhibits Gaussian statistics
with known covariance $N$. We observe that the generating function
$J$ is not needed, as we could equally well take derivatives with
respect to $j$ in order to generate moments.

Central to this partition function is the field prior

\begin{eqnarray}
\mathcal{P}(\varphi) & = & \mathcal{P}(\xi=G[\varphi])\ \left\Vert \frac{\delta G[\varphi]}{\delta\varphi'}\right\Vert \mathcal{P}(\varphi_{0})\label{eq:SupersymmetricProbability}\\
 & = & \underbrace{\frac{1}{\sqrt{\vert2\pi\Xi\vert}}e^{-\frac{1}{2}G[\varphi]^{\dagger}\Xi^{-1}G[\varphi]}}_{=\mathcal{:B}(\varphi)}\underbrace{\left\Vert \frac{\delta G[\varphi]}{\delta\varphi'}\right\Vert }_{=\mathcal{:J}(\varphi)}\mathcal{P}(\varphi_{0})
\end{eqnarray}
This contains a signal dependent term $\mathcal{B}(\varphi)$ from
the excitation statistics as well as another one, $\mathcal{J}(\varphi)$,
from the functional determinant. Especially the calculation of this
determinant remains a computational problem. The aim of the next section
is to represent the Jacobian determinant $\mathcal{J}$ by a path-integral
over fermionic fields for the data free partition function 
\begin{align}
\mathcal{Z} & =\int\mathcal{D}\varphi~\mathcal{P}(\varphi)=\int\mathcal{D}\varphi~e^{-\mathcal{H}(\varphi)}\nonumber \\
 & =\int\mathcal{D}\varphi~\mathcal{B}(\varphi)~\mathcal{J}(\varphi)\ \mathcal{P}(\varphi_{0}).
\end{align}

\section{Dynamical Field Inference with Ghost Fields}

\label{sec:Dynmical Field Inference with Ghost Fields}

\subsection{Grassmann fields}

\label{subsubsec:GrassmannVariables} Grassmann numbers $\{\chi_{1},\bar{\chi}_{1},\ldots\chi_{N},\bar{\chi}_{N}\}$
are independent elements, which anticommute among each other \citep{Wasay,QFT1,QFT2}
and thus follow the Pauli principle, $\chi_{i}^{2}=\bar{\chi}_{i}^{2}=0$
for $i\in\{1,\ldots N\}$. Consequently, a corresponding function
depending on the Grassmann numbers $\chi$ and $\bar{\chi}$ can be
Taylor expanded to 
\begin{align}
f(\chi,\bar{\chi})=a+b_{1}\chi+b_{2}\bar{\chi}+c_{12}\chi\bar{\chi}+c_{21}\bar{\chi}\chi.\label{eq:GrassmannExpansion}
\end{align}
A special feature of Grassmann numbers is that the integration and
differentiation to them are the same. As a consequence, one can write
down the following Grassmann integrals: 
\begin{align}
\int d\chi~d\bar{\chi} & =0\label{eq:1stGrassmannIntegral}\\
\int d\chi~d\bar{\chi}~\bar{\chi}\chi & =1\label{eq:2ndGrassmannIntegral}
\end{align}
In order to represent the Jacobian with infinite dimensions by a path
integral, we need to transform the Grassmann variables to Grassmann
fields with infinite dimensions. This leads us to path integrals over
Grassmann fields, 
\begin{align}
\int d\chi_{1}d\bar{\chi}_{1}~...~d\chi_{N}d\bar{\chi}_{N}\overset{N\to\infty}{\xrightarrow{\hspace*{0.8cm}}}\int\mathcal{D}\chi\mathcal{D}\bar{\chi},
\end{align}
with the following integration rules, 
\begin{align}
 & \int\mathcal{D}\chi~\mathcal{D}\bar{\chi}=0\\
 & \int\mathcal{D}\chi~\mathcal{D}\bar{\chi}~\bar{\chi}^{\dagger}\chi=\int\mathcal{D}\chi~\mathcal{D}\bar{\chi}~\biggl(\int_{\Omega'}\text{d}x\bar{\chi}(x)\chi(x)\biggr)=\mathds{1},
\end{align}
where the $\bar{\chi}^{\dagger}$ is the adjoint of the anti-commuting
field $\bar{\chi}$. The scalar product 
\begin{equation}
\bar{\chi}^{\dagger}\chi=\int_{\Omega'}\!\!\text{d}x\,\bar{\chi}(x)\chi(x)
\end{equation}
 will here be taken only over the domain $\Omega'$ without the inital
time slice, as the Grassmann fields are introduced to represent a
determinant of the functional $\mathcal{J}(\varphi)$, which is also
defined only over this domain. In the following, we abbreviate the
notation by writing $\int\!\text{d}x$ for $\int_{\Omega'}\text{d}x$.

\subsection{Path Integral Representation of Determinants and $\delta$-functions}

\label{subsubsec:Path Integral Representation of Determinants} By
means of the Grassmann fields we derive the path integral representation
for $\mathcal{J}$, the absolute value of the determinant of the Jacobian
$\frac{\delta G[\varphi]}{\delta\varphi'}$ \citep{Srednicki}. For
this purpose we take two unitary transformations $U$ and $V$ with
the property that $M=V\frac{\delta G[\varphi]}{\delta\varphi'}U$
becomes diagonal with positive and real entries. These are then used
to transform the Grassmann fields:
\begin{equation}
\chi=U\chi^{\prime},~~\bar{\chi}^{\dagger}=\bar{\chi}^{\prime\dagger}V
\end{equation}
This leads to a weighting of the path integral differentials by the
determinants of $U$ and $V$. 
\begin{equation}
\mathcal{D}\chi\mathcal{D}\bar{\chi}=\vert U\vert^{-1}\vert V\vert^{-1}\mathcal{D}\chi^{\prime}\mathcal{D}\bar{\chi}^{\prime}\label{eq:GrassmannTransformation}
\end{equation}
Here we used the identity of integration and differentiation for Grassmann
variables $\int d\chi=\frac{\partial}{\partial\chi}=\frac{\partial\chi^{\prime}}{\partial\chi}\frac{\partial}{\partial\chi^{\prime}}=|U|^{-1}\int d\chi^{\prime}$
to transform their differentials. The determinant of the operator
$M$ is given by the product of the operators, from which we can infer
the Jacobian determinant. 
\begin{align}
\vert M\vert & =\vert V\vert~\left|\frac{\delta G[\varphi]}{\delta\varphi'}\right|~\vert U\vert\\
\Rightarrow\left|\frac{\delta G[\varphi]}{\delta\varphi'}\right| & =\vert M\vert~\vert U\vert^{-1}~\vert V\vert^{-1}\label{eq:JacobianDeterminant}
\end{align}
As the operator $M$ is diagonal with eigenvalues $\{m_{i}\}$ on
the diagonal, we can write its determinant as a product of $N$ eigenvalues
in the limit of infinite dimensions $N$. 
\begin{align}
\vert M\vert & =\lim_{N\to\infty}\prod_{i=1}^{N}m_{i}\nonumber \\
 & \overset{(\ref{eq:1stGrassmannIntegral})}{\underset{(\ref{eq:2ndGrassmannIntegral})}{=}}\lim_{N\to\infty}\prod_{i=1}^{N}\left[\underbrace{\int d\chi_{i}^{\prime}~d\bar{\chi}_{i}^{\prime}}_{=0}+m_{i}\underbrace{\int d\chi_{i}^{\prime}~d\bar{\chi}_{i}^{\prime}~\bar{\chi}_{i}^{\prime}\chi_{i}^{\prime}}_{=1}\right]\nonumber \\
 & =\lim_{N\to\infty}\prod_{i=1}^{N}\int d\chi_{i}^{\prime}~d\bar{\chi}_{i}^{\prime}~(1+m_{i}\bar{\chi}_{i}^{\prime}\chi_{i}^{\prime})\nonumber \\
 & =\lim_{N\to\infty}\prod_{i=1}^{N}\int d\chi_{i}^{\prime}~d\bar{\chi}_{i}^{\prime}~(1+m_{i}\bar{\chi}_{i}^{\prime}\chi_{i}^{\prime}+\underbrace{\frac{1}{2!}m_{i}^{2}(\bar{\chi}_{i}^{\prime}\chi_{i}^{\prime})^{2}}_{\overset{(\ref{eq:GrassmannExpansion})}{=}0})\nonumber \\
 & =\lim_{N\to\infty}\prod_{i=1}^{N}\int d\chi_{i}^{\prime}~d\bar{\chi}_{i}^{\prime}~e^{m_{i}\bar{\chi}_{i}^{\prime}\chi_{i}^{\prime}}
\end{align}
The insertion of the result for the determinant of the diagonal matrix
$M$ in the definition of the Jacobian in Eq.~\eqref{eq:JacobianDeterminant}
yields 
\begin{align}
\left|\frac{\delta G[\varphi]}{\delta\varphi'}\right| & =\vert U\vert^{-1}~\vert V\vert^{-1}\lim_{N\to\infty}\prod_{i=1}^{N}\int d\chi_{i}^{\prime}~d\bar{\chi}_{i}^{\prime}~e^{m_{i}\bar{\chi}_{i}^{\prime}\chi_{i}^{\prime}}\nonumber \\
 & =\int\mathcal{D}\chi^{\prime}~\mathcal{D}\bar{\chi}^{\prime}~\vert U\vert^{-1}~\vert V\vert^{-1}e^{\bar{\chi}^{\prime\dagger}M\chi^{\prime}}\nonumber \\
 & =\int\mathcal{D}\chi^{\prime}~\mathcal{D}\bar{\chi}^{\prime}~\vert U\vert^{-1}~\vert V\vert^{-1}e^{\bar{\chi}^{\dagger}V^{-1}MU^{-1}\chi}\nonumber \\
 & \overset{(\ref{eq:GrassmannTransformation})}{=}\int\mathcal{D}\chi\mathcal{D}\bar{\chi}e^{\bar{\chi}^{\dagger}\frac{\delta G[\varphi]}{\delta\varphi'}\chi}.
\end{align}
Finally, we find the representation of the Jacobian in terms of an
integral over independent Grassmann fields, 
\begin{align}
\mathcal{J}=\left\Vert \frac{\delta G[\varphi]}{\delta\varphi^{\prime}}\right\Vert =\left|\int\mathcal{D}\chi\mathcal{D}\bar{\chi}~e^{\bar{\chi}^{\dagger}\frac{\delta G[\varphi]}{\delta\varphi'}\chi}\right|.\label{eq:FermionDeterminant}
\end{align}
We note that an equivalent expression is
\begin{align}
\mathcal{J}=\left\Vert -i\frac{\delta G[\varphi]}{\delta\varphi^{\prime}}\right\Vert =\left|\int\mathcal{D}\chi~\mathcal{D}\bar{\chi}~e^{-i\bar{\chi}^{\dagger}\frac{\delta G[\varphi]}{\delta\varphi^{\prime}}\chi}\right|,\label{eq:imaginary-unit}
\end{align}
as the factor $-i$ cancels out in taking the absolute value. In the
following, we will not track such multiplicative factors of unity
absolute value for probabilities, as these can be fixed at the end
of the calculation.

The other term in $\mathcal{P}(\varphi)=\mathcal{B}(\varphi)\,\mathcal{J}(\varphi)\,\mathcal{P}(\varphi_{0})$
as expressed by Eq.\ (\ref{eq:SupersymmetricProbability}), $\mathcal{B}(\varphi)=\mathcal{G}(G[\varphi],\Xi)$,
is highly non-Gaussian for a non-linear dynamics $G[\varphi]$. Here,
it is useful to step back to the initial form including the excitation
field
\begin{equation}
\mathcal{B}(\varphi)=\int\mathcal{D}\xi~\delta(\xi-G[\varphi])\,e^{-\mathcal{H}(\xi)},
\end{equation}
with $\mathcal{H}(\xi)=-\ln\mathcal{G}(\xi,\Xi)=\frac{1}{2}\xi^{\dagger}\Xi^{-1}\xi+\frac{1}{2}\ln\vert2\pi\Xi\vert$,
and to replace the $\delta$-function by means of a path integral.
In order to do so the representation of the $\delta$-function as
an integral over Fourier modes is recalled. 
\begin{align}
\delta(x)=\frac{1}{2\pi}\int dk~e^{-ikx}
\end{align}
The migration of this to path-integral representation is achieved
by the introduction of a Lagrange multiplier field $\beta(x)$, 
\begin{align}
\delta(\xi) & =\frac{1}{\left|2\pi\mathbb{1}\right|}\int\mathcal{D}\beta~e^{-i\beta^{\dagger}\xi}.\label{eq:BosonDelta}
\end{align}
With this, the field prior reads
\begin{eqnarray}
\mathcal{P}(\varphi) & \propto & \int\frac{\mathcal{D}\xi\mathcal{D}\beta\mathcal{D}\chi\mathcal{D}\bar{\chi}}{\sqrt{\vert2\pi\Xi\vert}\left|2\pi\mathbb{1}\right|}\ e^{-\frac{1}{2}\xi^{\dagger}\Xi^{-1}\xi-\mathcal{H}(\varphi_{0})}\ \times\nonumber \\
 &  & e^{-i\bigl(\bar{\chi}^{\dagger}\frac{\delta G[\varphi]}{\delta\varphi'}\chi-\beta^{\dagger}(G[\varphi]-\xi)\bigr)}
\end{eqnarray}
with $\mathcal{H}(\varphi_{0})=-\ln\mathcal{P}(\varphi_{0})$ the
information on the initial conditions.

\subsection{Ghost Field Path Integrals in DFI}

\label{subsec:Ghost Field Path Integrals in DFI} With the introduction
of the fields $\beta$, $\chi$, and $\bar{\chi}$ the \ac{DFI} partition
function is now given by path integrals over the excitations and additional
two fermionic and two bosonic degrees of freedom, which are summarized
to a tuple of fields $\psi=(\varphi,~\beta,~\chi,~\bar{\chi})$,\footnote{Note, the here defined $\psi$ differs from the initially introduced
system state, also denoted by $\psi$. As the latter will not be used
any more in this work, the reuse of the symbol is hopefully acceptable.} 
\begin{align}
\mathcal{Z}\propto\int\mathcal{D}\xi~\mathcal{D}\psi~e^{-\mathcal{H}(\xi)-\mathcal{H}(\varphi_{0})+i\beta^{\dagger}(G[\varphi]-\xi)-i\bar{\chi}^{\dagger}\frac{\delta G[\varphi]}{\delta\varphi'}\chi}.\label{eq:2Boson2FermionPartitionFunction}
\end{align}
\textbf{ }Let us introduce the functional $\{Q[\psi],\cdot\}=\{Q[\chi,\beta],\cdot\}$,
which depends on the fermionic ghost field $\chi$ and the bosonic
Lagrange multiplier $\beta$ 
\begin{align}
\{Q,X\}[\psi] & =\int\text{d}x~\biggl(\beta(x)\frac{\delta}{\delta\bar{\chi}(x)}+\chi(x)\frac{\delta}{\delta\varphi^{\prime}(x)}\biggr)X[\psi]\nonumber \\
 & =\biggl(\beta\frac{\delta}{\delta\bar{\chi}}+\chi\frac{\delta}{\delta\varphi^{\prime}}\biggr)^{T}X[\psi].\label{eq:QP-Bracket}
\end{align}

Next, the exponent of the partition function in Eq.~\eqref{eq:2Boson2FermionPartitionFunction}
is reshaped in order to be $Q$-exact. This means that the exponent
shall only depend on the introduced functional $\{Q,\cdot\}$ for
a suitable $X$. For this we investigate the two ghost and Lagrange
field dependent terms in Eq.~\eqref{eq:2Boson2FermionPartitionFunction}
separately.

The fermionic ghost field dependent exponent is 
\begin{align}
E_{\text{fg}} & =-i\bar{\chi}^{\dagger}\frac{\delta G[\varphi]}{\delta\varphi^{\prime}}\chi\nonumber \\
 & =-i\int\text{d}x'\text{d}x~\bar{\chi}(x)\frac{\delta G[\varphi](x)}{\delta\varphi^{\prime}(x')}\chi(x')\nonumber \\
 & =i\int\text{d}x'\text{d}x~\chi(x')\frac{\delta}{\delta\varphi^{\prime}(x')}G[\varphi](x)\bar{\chi}(x)\nonumber \\
 & =i\biggl(\chi^{\dagger}\frac{\delta}{\delta\varphi^{\prime}}\biggr)\biggl(\bar{\chi}^{\dagger}~G[\varphi]\biggr)\nonumber \\
 & =i\biggl(\chi^{\dagger}\frac{\delta}{\delta\varphi^{\prime}}\biggr)\biggl(\bar{\chi}^{\dagger}~(G[\varphi]-\xi)\biggr)\label{eq:ProofSUSYFunct1}
\end{align}
and the bosonic Lagrange field dependent exponent is 
\begin{align}
E_{\text{bg}} & =i\beta^{\dagger}(G[\varphi]-\xi)\nonumber \\
 & =i\int\text{d}x~\beta(x)(G[\varphi](x)-\xi(x))\nonumber \\
 & =i\int\text{d}x~\text{d}x'~\beta(x')\frac{\delta\bar{\chi}(x)}{\delta\bar{\chi}(x')}(G[\varphi](x)-\xi(x))\nonumber \\
 & =i\biggl(\beta^{T}\frac{\delta}{\delta\bar{\chi}}\biggr)\biggl(\bar{\chi}^{\dagger}~(G[\varphi]-\xi)\biggr).\label{eq:ProofSUSYFunct2}
\end{align}
Thus the whole ghost and Lagrange field dependent exponent can be
written as a $Q$-exact expression: 
\begin{eqnarray}
E_{\text{fg}}+E_{\text{bg}} & = & i\beta^{\dagger}(G[\varphi]-\xi)-i\bar{\chi}^{\dagger}\frac{\delta G[\varphi]}{\delta\varphi^{\prime}}\chi\nonumber \\
 & \underset{(\ref{eq:ProofSUSYFunct2})}{\overset{(\ref{eq:ProofSUSYFunct1})}{=}} & i\biggl(\chi^{\dagger}~\frac{\delta}{\delta\varphi^{\prime}}+\beta^{\dagger}~\frac{\delta}{\delta\bar{\chi}}\biggr)\biggl(\bar{\chi}^{\dagger}~(G[\varphi]-\xi)\biggr)\nonumber \\
 & = & i\{Q,\bar{\chi}^{\dagger}(G[\varphi]-\xi)\}
\end{eqnarray}

According to these auxiliary calculations, the partition function
in Eq.~\eqref{eq:2Boson2FermionPartitionFunction} takes the form,
\begin{align}
\mathcal{Z} & \propto\int\mathcal{D}\xi~\mathcal{D}\psi~e^{-\mathcal{H}(\xi)-\mathcal{H}(\varphi_{0})+i\{Q,\bar{\chi}^{\dagger}(G[\varphi]-\xi)\}}.\label{eq:NoiseSupersymmetric PartitionFunction}
\end{align}
The integration over the excitation fields creates a partition function
that only contains the fields of the set $\psi=(\varphi,~\beta,~\chi,~\bar{\chi})$.
With the aid of the following relation for a bosonic field $y(x)$
that is independent of $\varphi$ 
\begin{align}
\{Q,\bar{\chi}^{\dagger}y\} & =\biggl(\beta^{\dagger}\frac{\delta}{\delta\bar{\chi}}\biggr)\bar{\chi}^{\dagger}y\nonumber \\
 & =\int\text{d}x'~\beta(x')\frac{\delta}{\delta\bar{\chi}(x')}\int\text{d}x~\bar{\chi}(x)y(x)\nonumber \\
 & =\int\text{d}x~\beta(x)y(x)\nonumber \\
 & =\beta^{\dagger}y\label{eq:QExactGaussian}
\end{align}
the integration over the excitation field can be performed for a Gaussian
excitation field ($\mathcal{H}(\xi)\,\widehat{=}\,\frac{1}{2}\xi^{\dagger}\xi$):
\begin{align}
\mathcal{Z} & ~\propto\int\mathcal{D}\psi~\mathcal{D}\xi~e^{i\{Q,\bar{\chi}^{\dagger}G[\varphi]\}-i\{Q,\bar{\chi}^{\dagger}\xi\}-\mathcal{H}(\xi)-\mathcal{H}(\varphi_{0})}\nonumber \\
 & \overset{(\ref{eq:QExactGaussian})}{=}\int\mathcal{D}\psi~\mathcal{D}\xi~e^{i\{Q,\bar{\chi}^{\dagger}G[\varphi]\}-i\beta^{\dagger}\xi-\mathcal{H}(\xi)-\mathcal{H}(\varphi_{0})}\nonumber \\
 & ~=\int\mathcal{D}\psi~\mathcal{D}\xi~e^{i\{Q,\bar{\chi}^{\dagger}G[\varphi]\}-i\beta^{\dagger}\xi-\frac{1}{2}\xi^{\dagger}\Xi^{-1}\xi-\mathcal{H}(\varphi_{0})}\nonumber \\
 & ~=\int\mathcal{D}\psi~e^{i\{Q,\bar{\chi}^{\dagger}G[\varphi]\}-\frac{1}{2}\beta^{\dagger}\Xi\beta-\mathcal{H}(\varphi_{0})}\nonumber \\
 & ~=\int\mathcal{D}\psi~e^{i\{Q,\bar{\chi}^{\dagger}G[\varphi]\}-\frac{1}{2}\{Q,\bar{\chi}^{\dagger}\Xi\beta\}-\mathcal{H}(\varphi_{0})}\nonumber \\
 & ~=\int\mathcal{D}\psi~e^{\{Q,i\bar{\chi}^{\dagger}G[\varphi]-\frac{1}{2}\bar{\chi}^{\dagger}\Xi\beta\}-\mathcal{H}(\varphi_{0})}
\end{align}
 Now, we define the odd function 
\begin{align}
\theta(\psi) & =\bar{\chi}^{\dagger}(-iG[\varphi]+\frac{1}{2}\Xi\beta)\label{eq:GaugeFermion}
\end{align}
for reasons of clarity. Besides we revive the statistical mechanics
formalism for the definition of the partition function from Eq.~\eqref{eq:GeneralPartitionFunction}
as well as the corresponding ghost and Lagrange field dependent information
$\mathcal{H}(\psi)$: 
\begin{align}
\mathcal{Z} & =\int\mathcal{D}\psi~e^{-\mathcal{H}(\varphi_{0})-\mathcal{H}(\psi|\varphi_{0})}\label{eq:QExactPartitionFunction1}\\
 & \propto\int\mathcal{D}\psi~e^{-\mathcal{H}(\varphi_{0})-\{Q,\theta(\psi)\}}\label{eq:QExactPartitionFunction2}\\
\mathcal{H}(\psi|\varphi_{0}) & \,\widehat{=}\,\{Q,\theta(\psi)\}\label{eq:QExactHamiltonian}
\end{align}
Here, $\widehat{=}$ indicates equality up to a constant term due
to the not tracked absolute phase of our expressions. By comparison
we find the following relation between the prior information Hamiltonian
of the signal field $\mathcal{H}(\varphi)$ and the just derived information
Hamiltonian of the ghost and Lagrange fields. 
\begin{align}
\mathcal{Z} & \overset{(\ref{eq:GeneralPartitionFunction})}{\propto}\int\mathcal{D}\varphi~e^{-\mathcal{H}(\varphi|\varphi_{0})-\mathcal{H}(\varphi_{0})}\\
 & \overset{(\ref{eq:QExactPartitionFunction1})}{=}\int\mathcal{D}\psi~e^{-\mathcal{H}(\psi|\varphi_{0})-\mathcal{H}(\varphi_{0})}\\
\Rightarrow & \mathcal{H}(\varphi|\varphi_{0})=-\ln\biggl(\int\mathcal{D}\chi~\mathcal{D}\bar{\chi}~\mathcal{D}\beta~e^{-\mathcal{H}(\psi|\varphi_{0})}\biggr)\label{eq:IFT-STS-connection}
\end{align}
Let us now emphasize the first time derivative in the \ac{SDE} by
taking the definition of the \ac{SDE} from Eq.~\eqref{eq:TimeDep.SDE},
$F[\varphi^{\prime}](x)+\xi(x)=\partial_{t}\varphi(x)$, so that the
$\theta$-functional becomes 
\begin{align}
\theta(\psi) & =+i\bar{\chi}^{\dagger}F[\varphi^{\prime}]-i\bar{\chi}^{\dagger}\partial_{t}\varphi+\frac{1}{2}\bar{\chi}^{\dagger}\Xi\beta\nonumber \\
 & =-i\bar{\chi}^{\dagger}\partial_{t}\varphi+i\bar{Q}(\psi).
\end{align}
Here we introduced the functional on the set of fields $\psi$ 
 
\begin{align}
\bar{Q}(\psi)=\bar{\chi}^{\dagger}F[\varphi^{\prime}]-i\frac{1}{2}\bar{\chi}^{\dagger}\Xi\beta.\label{eq:QFunctional}
\end{align}
Evaluating the information for this $\theta$-functional one gets
\begin{align}
\mathcal{H}(\psi|\varphi_{0}) & \overset{(\ref{eq:QExactHamiltonian})}{\widehat{=}}-\{Q,i\bar{\chi}^{\dagger}\partial_{t}\varphi\}+i\{Q,\bar{Q}\}.\label{eq:HQQ}
\end{align}

The Fermionic field $\chi$ was only defined over $\Omega'$ the field
domain without the initial time slice in order to represent the determinant
of the Jacobian of $G(\varphi)$ with respect to $\varphi'.$ One
can extend the support of $\chi$ to $\Omega$, including the initial
time slice by introducing a split notation for this extended $\chi=(\chi_{0},\chi^{\prime})^{\dagger}$,
with $\chi'$ denoting the original Fermionic field over $\Omega'$.
We then find that the ghost field has to vanish at the initial time
step $t_{0}$, i.e. $\chi=(0,\chi^{\prime})^{\dagger}$ in order to
assure that the following expression does not diverge. Here, we abbreviate
$\varphi_{t}=\varphi(x)=\varphi(\vec{x},t)\:\text{and }\varphi_{t+\Delta}=\varphi(\vec{x},t+\Delta t)$:

\begin{align}
 & \{Q,i\bar{\chi}^{\dagger}\partial_{t}\varphi\}\nonumber \\
= & \bigg(\chi^{\dagger}\frac{\delta}{\delta\varphi^{\prime}}+\beta^{\dagger}\frac{\delta}{\delta\bar{\chi}}\bigg)\ i\bar{\chi}^{\dagger}\partial_{t}\varphi\nonumber \\
= & \underbrace{i\chi^{\dagger}\frac{\delta}{\delta\varphi^{\prime}}\bar{\chi}^{\dagger}\partial_{t}\varphi}_{=A}+\underbrace{i\beta^{\dagger}\frac{\delta}{\delta\bar{\chi}}\bar{\chi}^{\dagger}\partial_{t}\varphi}_{=B}\\
A= & i\int\text{d}x'\,\text{d}x\,\chi(x')\,\chi(x)\frac{\delta}{\delta\varphi'(x')}\partial_{t}\varphi(x)\nonumber \\
= & -i\int\text{d}x\,\text{d}x'\bar{\chi}(x)\,\chi(x')\frac{\delta}{\delta\varphi'(x')}\lim_{\Delta t\to0}\biggl(\begin{array}{c}
\frac{\varphi_{0+\Delta}-\varphi_{0}}{\Delta t}\\
\frac{\varphi_{t+\Delta}-\varphi_{t}}{\Delta t}
\end{array}\biggr)\nonumber \\
= & -i\int\text{d}x\,\text{d}x'\bar{\chi}(x)\,\chi(x')\,\lim_{\Delta t\to0}\biggl(\begin{array}{c}
\frac{\delta_{\vec{x},\vec{x}^{\prime}}\delta_{0+\Delta,t^{\prime}}}{\Delta t}\\
\frac{\delta_{\vec{x},\vec{x}^{\prime}}(\delta_{t+\Delta,t^{\prime}}-\delta_{t,t^{\prime}})}{\Delta t}
\end{array}\biggr)\nonumber \\
= & -i\int\text{d}x\bar{\chi}(x)\,\lim_{\Delta t\to0}\left(\begin{array}{c}
\frac{\chi_{_{0+\Delta}}}{\Delta t}\\
\frac{\chi_{t+\Delta}-\chi_{t}}{\Delta t}
\end{array}\right)\nonumber \\
= & -i\int\text{d}x\bar{\chi}(x)\,\left(\begin{array}{c}
\begin{cases}
0 & \text{ if }\chi_{0}=0\\
\infty & \text{otherwise}
\end{cases}\\
\partial_{t}\chi'
\end{array}\right)\label{eq:crucial_step}\\
= & -i\int\text{d}x\bar{\chi}(x)\,\partial_{t}\chi(x)\label{eq:after-crucial_step}\\
= & -i\bar{\chi}^{\dagger}\partial_{t}\chi\\
B= & i\!\int\!\text{d}x'\beta(x')\!\int\!\text{d}x~\frac{\delta\bar{\chi}(x)}{\delta\bar{\chi}(x')}\partial_{t}\varphi(x)\nonumber \\
= & i\!\int\!\text{d}x'\beta(x')\!\int\!\text{d}x~\delta(x-x')\partial_{t}\varphi^{\prime}(x)\nonumber \\
= & i\!\int\!\text{d}x\!\int\!\text{d}x'\ \beta(x')\delta(x-x')\partial_{t}\varphi(x)\nonumber \\
= & i\!\int\!\text{d}x~\beta(x)\partial_{t}\varphi(x)\nonumber \\
= & i\beta^{\dagger}\partial_{t}\varphi
\end{align}
 such that, 
\begin{align}
\mathcal{H}(\psi|\varphi_{0})\, & \widehat{=}\,i\bar{\chi}^{\dagger}\partial_{t}\chi-i\beta^{\dagger}\partial_{t}\varphi+i\{Q,\bar{Q}\}.\label{eq:SupersymmetricIFTHamiltonian}
\end{align}

The crucial insight is given by Eq.\ \eqref{eq:crucial_step}. If
$\chi_{0}\neq0,$the expression $A$ would diverge and Eq.\ \eqref{eq:HQQ}
would not hold. In order to reestablish a compact notation in\emph{
}Eq.\ \eqref{eq:after-crucial_step}, we note that any finite assignment
of $\partial_{t}\chi_{0}\neq0$ would only make a vanishing contribution
to the integral as being on an infinitesimal smart support.

The information Hamiltonian of Eq.\ \eqref{eq:HQQ} has two parts.
We call the left part, which contains the time derivatives of the
fermionic and bosonic fields, the dynamic information. The right part,
which is described by the Poisson bracket, is referred to as the static
information. The derivation of Poisson brackets in a system with fermionic
and bosonic fields is described in \citep{OddPoisson,SUSYClassical}.

This yields the partition function, 
\begin{align}
\mathcal{Z}\propto\int\mathcal{D}\psi~e^{-i\bar{\chi}^{\dagger}\partial_{t}\chi+i\beta^{\dagger}\partial_{t}\varphi-i\{Q,\bar{Q}\}-\mathcal{H}(\varphi_{0})}.\label{eq:IFTPartitionFunction}
\end{align}
So far we represented the partition function in terms of the signal
field, $\varphi$, and the three fields, $\beta,~\chi,~\bar{\chi}$.

In case of a white excitation field $\xi$ the partition function
of \ac{DFI} can be derived using the Markov property. For this, we
start with the \ac{IFT} partition function for a bosonic field $\varphi$
and a fermionic field $\chi$ and decompose it in terms of time-ordered
conditional probabilities
\begin{align}
 & \mathcal{Z}=\int\mathcal{D}\varphi~\mathcal{D}\chi~\mathcal{P}(\varphi,\chi)\\
 & =\left(\prod_{n=0}^{N}\int\mathcal{D}\varphi_{n}~\int\mathcal{D}\chi_{n}\right)~\mathcal{P}(\varphi_{N},\chi_{N},\varphi_{N-1},\chi_{N-1},\nonumber \\
 & ~~~\varphi_{N-2},...,\varphi_{1},\chi_{1},\varphi_{0}),\\
\text{} & =\left(\prod_{n=0}^{N}\int\mathcal{D}\varphi_{n}~\int\mathcal{D}\chi_{n}\right)~\mathcal{P}(\varphi_{N},\chi_{N}|\varphi_{N-1},\chi_{N-1})\nonumber \\
 & ~~\times.......\times\mathcal{P}(\varphi_{1},\chi_{1}|\varphi_{0})\mathcal{P}(\varphi_{0})\label{eq:Discomposed PartitionFunction}
\end{align}
where $\varphi_{0}=\varphi(\cdot,t_{0})$ is the field at initial
time $t_{0}=0$ while there is no $\chi_{0}=\chi(\cdot,t_{0})$ .

The conditional probabilities can then be represented as QFT transition
amplitudes \citep{TransitionAmplitude1,TransitionAmplitude2} between
states of the system denoted by the Dirac notation as 
\begin{eqnarray}
\mathcal{P}(\varphi_{k},\,\chi_{k}|\varphi_{j},\,\chi_{j}) & =: & \text{\ensuremath{\braket{\varphi_{k},\chi_{k},t_{k}||\varphi_{j},\chi_{j},t_{j}}}}\nonumber \\
 & := & \bra{\varphi_{k},\chi_{k}}\mathcal{M}(t_{k},t_{j})\ket{\varphi_{j},\chi_{j}}
\end{eqnarray}
At this stage, these are formal definitions, with the time localized
states $\bra{\varphi_{k},\chi_{k},t_{k}}:=\delta(\varphi(\cdot,t_{k})-\varphi_{k})\,\delta(\chi(\cdot,t_{k})-\chi_{k})$,
$\ket{\varphi_{j},\chi_{j},t_{j}}:=\delta(\varphi(\cdot,t_{j})-\varphi_{j})\,\delta(\chi(\cdot,t_{j})-\chi_{j})$,
and the not localized ones $\bra{\varphi_{k},\chi_{k}}:=\delta(\varphi(\cdot,t)-\varphi_{k})\,\delta(\chi(\cdot,t)-\chi_{k})$,
$\ket{\varphi_{j},\chi_{j}}:=\delta(\varphi(\cdot,t)-\varphi_{j})\,\delta(\chi(\cdot,t)-\chi_{j})$,
with $t$ being some unspecified time. Here, $j$ and $k$ label time-slice
field configurations, like $\varphi(\cdot,t)=\varphi_{j}$ and $\varphi(\cdot,t)=\varphi_{k}$,
and their associated times are $t=t_{j}$ and $t=t_{k}.$ The first
line does not contain a usual scalar product between states, as the
variables have first to be brought to a common time. This is done
in the second line by the transfer operator $\mathcal{M}(t_{k},t_{j})$,
which describes the mapping of states at time $t_{j}$ to such at
$t_{k}$. In \citep{Ovchinnikov2} it is shown that a representation
of these state vectors is given by the exterior algebra over the field
configuration space.

By assigning field operators to the fermionic and bosonic fields,
$\chi$ and $\varphi$, as well as their momenta, $\nu$ and $\omega$
respectively, the partition function in Eq.\ \eqref{eq:Discomposed PartitionFunction}
can be rewritten in terms of the generalized Fokker-Planck operator
of the states $\hat{H}$ according to \citep{PathIntegrals,TransitionAmplitude1,TransitionAmplitude2,QM}.
$\hat{H}$ is not to be confused with the information Hamiltonian
$\mathcal{H}(\psi|\varphi_{0})$. The precise relation of these will
be established in the following.

As mentioned in \citep{Ovchinnikov1,Ovchinnikov2,Ovchinnikov3,Ovchinnikov4},
the time evolution operator $\hat{H}$ is not Hermitian and thus the
time evolution is not described by the Schrödinger equation but by
the generalized Fokker-Planck equation instead.
\begin{eqnarray}
\partial_{t}\ket{\varphi,\chi,t} & = & -\hat{H}\ket{\varphi,\chi,t}\label{eq:FPE}\\
\Rightarrow\ket{\varphi,\chi,t+\Delta t} & = & e^{-\hat{H}\Delta t}\ket{\varphi,\chi,t}\\
\Rightarrow\mathcal{M}(t_{k},t_{j}) & = & e^{-\hat{H}(t_{k}-t_{j})}
\end{eqnarray}
These and the following equations define the properties of Ĥ. The
conditional probabilities for the fields $\varphi_{k}$ and $\chi_{k}$
given the fields at the previous time step $\varphi_{k-1}$, $\chi_{k-1}$
are given by the transition amplitudes between the corresponding states
and are defined via the time evolution. 
\begin{eqnarray}
\mathcal{P}_{k,k-1} & = & \mathcal{P}(\varphi_{k},\chi_{k}|\varphi_{k-1},\chi_{k-1})\nonumber \\
 & = & \text{\ensuremath{\braket{\varphi_{k},\chi_{k},t_{k}||\varphi_{k-1},\chi_{k-1},t_{k-1}}}}\nonumber \\
 & = & \braket{\varphi_{k},\chi_{k}|e^{-\hat{H}\Delta t}|\varphi_{k-1},\chi_{k-1}}
\end{eqnarray}
At this point we multiply with unity, 
\begin{eqnarray}
\mathbb{1} & = & \int\mathcal{D}\omega_{k}~\mathcal{D}\nu_{k}\,\ket{\omega_{k},\nu_{k}}\bra{\omega_{k},\nu_{k}},
\end{eqnarray}
where the $\ket{\omega_{k},\nu_{k}}$ are momentum eigenstates of
the field that obey on equal time slices
\begin{eqnarray}
\braket{\omega_{k},\nu_{k}|\varphi_{k},\chi_{k}} & = & e^{-i\omega_{k}\varphi_{k}+i\nu_{k}\chi_{k}}.
\end{eqnarray}
If we choose infinitesimal small time steps, we can evaluate the time-evolution
operator on the momentum eigenstate, which leads to the following
expression for the conditional probability
\begin{eqnarray}
\mathcal{P}_{k,k-1} & = & \int\mathcal{D}\omega_{k}~\mathcal{D}\nu_{k}\,\bra{\varphi_{k},\chi_{k}}e^{-\hat{H}\Delta t}\nonumber \\
 &  & \times\ket{\omega_{k},\nu_{k}}\braket{\omega_{k},\nu_{k}|\varphi_{k-1},\chi_{k-1}}\nonumber \\
 & = & \int\mathcal{D}\omega_{k}~\mathcal{D}\nu_{k}\,e^{-i\omega_{k}\varphi_{k-1}+i\nu_{k}\chi_{k-1}}\nonumber \\
 &  & \times\braket{\varphi_{k},\chi_{k}|e^{-\hat{H}\Delta t}|\omega_{k},\nu_{k}}\nonumber \\
 & \propto & \int\mathcal{D}\omega_{k}~\mathcal{D}\nu_{k}~e^{-H(\varphi_{k},\chi_{k},\omega_{k},\nu_{k})\Delta t-i\omega_{k}\varphi_{k-1}}\nonumber \\
 &  & \times e^{+i\nu_{k}\chi_{k-1}}\braket{\varphi_{k},\chi_{k}|\omega_{k},\nu_{k}}\nonumber \\
 & = & \int\mathcal{D}\omega_{k}\mathcal{D}\nu_{k}e^{-H(\varphi_{k},\chi_{k},\omega_{k},\nu_{k})\Delta t+i\omega_{k}(\varphi_{k}-\varphi_{k-1})}\nonumber \\
 &  & \times e^{-i\nu_{k}(\chi_{k}-\chi_{k-1})},
\end{eqnarray}
 The formal definition of $H(\varphi_{k},\chi_{k},\omega_{k},\nu_{k})$
for $\Delta t\to0$ is:
\begin{equation}
H(\varphi_{k},\chi_{k},\omega_{k},\nu_{k})=-\frac{1}{\Delta t}\ln\braket{\varphi_{k},\chi_{k}|e^{-\hat{H}\Delta t}|\omega_{k},\nu_{k}}\label{eq:Time-evolution}
\end{equation}
With this in mind the conditional transition probability distributions
can be written in terms of the function $H$. In the next step these
are inserted into the partition function in Eq.~\eqref{eq:Discomposed PartitionFunction}.
Taking the limit $\Delta t\to0$, $N\to\infty$ leads to

\begin{align}
\mathcal{Z} & \text{\ensuremath{\propto}}\int\mathcal{D}\psi~e^{-\int\text{d}t\,H(\varphi_{t},\chi_{t},\omega_{t},\nu_{t})+i\omega^{\dagger}\partial_{t}\varphi-i\nu^{\dagger}\partial_{t}\chi-\mathcal{H}(\varphi_{0})}.\label{eq:QFTPartitionFunction}
\end{align}

In the end, the partition function in Eq.~\eqref{eq:IFTPartitionFunction}
needs to be equal to the partition function in Eq.~\eqref{eq:QFTPartitionFunction}
in order to guarantee consistency of the theory. This permits the
following identifications, 
\begin{align}
\nu & =\bar{\chi},\\
\omega & =\beta,\\
\int\text{d}t\,H(\psi_{t}) & =i\{Q(\psi),\bar{Q}(\psi)\}.\label{eq:SUSYHamiltonian}
\end{align}
To sum up, it was shown that the auxiliary fields $\bar{\chi}$ and
$\beta$ are simply the momenta of the ghost field $\chi$ and the
signal field $\varphi$, respectively. And, for the moment the more
important finding is that the time evolution is governed by the $Q$-exact
static information , i.e. $\int\text{d}t\,H(t)=i\{Q,\bar{Q}\}$. Comparing
Eq.\ \eqref{eq:SupersymmetricIFTHamiltonian} to Eq.\ \eqref{eq:SUSYHamiltonian},
we find this enters directly the information Hamiltonian,
\begin{equation}
\mathcal{H}(\psi|\varphi_{0})\,\widehat{=}\,i\bar{\chi}^{\dagger}\partial_{t}\chi-i\beta^{\dagger}\partial_{t}\varphi+\int\text{d}t\,H(\psi_{t}),
\end{equation}
which can be regarded in combination with Eq.\ \eqref{eq:IFT-STS-connection}
as the central connection between STS and IFT, relating the information
Hamiltonian $\mathcal{H}(\psi|\varphi_{0})$ for the full system trajectory
to the Fokker-Planck evolution operators $H(\psi_{t})$ on individual
time-slices. $\mathcal{H}$ is a dimensionless quantity, whereas $H$
has the units of a rate.

In \citep{Ovchinnikov5} it is shown that $\{Q,\cdot\}$ is the path-integral
version of the exterior derivative $\hat{d}$ in the exterior algebra.
This recognition allows to identify the time-evolution in Eq.~(\ref{eq:SUSYHamiltonian})
as the path-integral version of the time-evolution operator in the
Focker-Planck equation. Moreover it is demonstrated that this time-evolution
operator is $\hat{d}$-exact and since the exterior derivative is
nilpotent the exterior derivative commutes with the time-evolution.
The conlusion is made that this corresponds to a supersymmetry. Firstly,
$\hat{d}$ as the operator representative of $\{Q,\cdot\}$ interchanges
fermions and bosons, since it replaces one bosonic field variable
by a fermionic one. Secondly, since a physical system is symmetric
with regard to an operator, if the operator commutes with the time-evolution
operator. As this is the case for $\hat{d}$ and $\hat{H}$, the field
dynamics is supersymmetric.

\subsection{Spontaneous SUSY Breaking and Field Inference}

The supersymetry of a dynamical field can be spontaneously broken
\citep{Ovchinnikov1,Ovchinnikov2,Ovchinnikov3,Ovchinnikov4}. This
coincides with the appearance of dynamical chaos as characterized
by positive Lyaponov exponents for the growth of the difference of
nearby system trajectories. It is intuitively clear that the occurrence
of chaos will reduce the predictability of the system and therefore
make field inference from measurements more difficult. We hope that
the here established connection of DFI and STS will permit to quantify
the impact of chaos on field inference in future research. For the
time being, we investigate the reverse impact, that of measurements
on the supersymmetry of the field knowlege as encoded in the partition
function.

\section{SUSY and Measurements}

\subsection{Abstract Considerations}

\label{sec:SUSY-Breaking-by}In Sec.\ \ref{subsec:Partition-Function}
we introduced the moment generating function in IFT in order to calculate
field expectation values after measurement data $d$ became available.
For a dynamical field, this can now be written with the help of STS
according to Eq.~\eqref{eq:momentGeneratingFunction} as
\begin{eqnarray}
\mathcal{Z}_{d}[J] & \!=\! & \int\mathcal{D}\varphi~e^{-\mathcal{H}(d,\varphi)+J^{\dagger}\varphi}\nonumber \\
 & \!=\! & \int\mathcal{D}\varphi\mathcal{P}(\varphi)\mathcal{P}(d|\varphi)e^{J^{\dagger}\varphi}\nonumber \\
 & \!\overset{(\ref{eq:SupersymmetricProbability})}{=}\! & \int\mathcal{D}\varphi\left\Vert \frac{\delta G[\varphi]}{\delta\varphi^{\prime}}\right\Vert \mathcal{G}(G[\varphi],\Xi)\mathcal{P}(\varphi_{0})\mathcal{P}(d|\varphi)e^{J^{\dagger\varphi}}\nonumber \\
 & \!\propto\! & \int\mathcal{D}\psi e^{\{Q,-\bar{\chi}^{\dagger}G[\varphi]-\frac{1}{2}\bar{\chi}^{\dagger}\Xi\beta\}-\mathcal{H}(\varphi_{0})-\mathcal{H}(d|\varphi)+J^{\dagger}\varphi}.\nonumber \\
\label{eq:final-partition function}
\end{eqnarray}
Note that we removed the $-i$ factor from the Fermionic variables
that was introduced in Eq.\ \ref{eq:imaginary-unit} in order to
connect to the conventions of the STS literature. Doing so, alleviates
us from the necessity to take the absolute value from the corresponding
term. From Eq.\ \ref{eq:final-partition function} we see that the
combined information representing the knowledge from measurement data
$d$ and about the dynamics as expressed by the $\theta$-function
from Eq.~\eqref{eq:GaugeFermion} consists of several parts,
\begin{eqnarray}
\mathcal{H}(d,\psi) & \widehat{=} & \{Q,\theta(\psi)\}+\mathcal{H}(d|\varphi)+\mathcal{H}(\varphi_{0})\nonumber \\
 & = & -\bar{\chi}^{\dagger}\partial_{t}\chi-i\beta^{\dagger}\partial_{t}\varphi^{\prime}+\{Q(\psi),\bar{Q}(\psi)\}\nonumber \\
 &  & +\mathcal{H}(d|\varphi)+\mathcal{H}(\varphi_{0}).\label{eq:total-information-source}
\end{eqnarray}
The first part, $-\bar{\chi}^{\dagger}\partial_{t}\chi-i\beta^{\dagger}\partial_{t}\varphi^{\prime}+\{Q(\psi),\bar{Q}(\psi)\},$
describes the dynamics of the field $\varphi'$ and that of the ghost
fields $\chi$ and $\overline{\chi}$ for times after the initial
moment by a $Q$-exact term, meaning that supersymmetry is conserved
if only this would affect the fields for non-inital times $t>t_{0}$.
The last term, $\mathcal{H}(\varphi_{0})=-\ln\mathcal{P}(\varphi_{0})$,
describes our knowledge on the initial conditions and not of the evolving
field. The middle term, $\mathcal{H}(d|\varphi)=-\ln\mathcal{P}(d|\varphi)$,
describes the knowledge gain by the measurement. If it addresses non-inital
times, it is in general not $Q$-exact. Thus, if one would take the
perspective of including the measurement constraints into the system
dynamics, as it was done with the noise excitation, the thereby extended
system would not be $Q$-exact any more. The reason for this is that
``external forces'' need to be introduced into the system description
to guide its evolution through the constraints set by the measurement,
which are not stationary and Gaussian as the excitation noise is.
Or more precisely, the knowledge state on the excitation field $\xi$
is in general not a zero-centered Gaussian prior with a stationary
correlation structure any more, but a posterior $\mathcal{P}(\xi|d)$
with explicitly time-dependent mean and correlation structure in $\xi$.

\subsection{Idealized Linear Dynamics\label{subsec:Idealized-Linear-Dynamics}}

In order to illustrate the impact of chaos on the predictability of
a system, we analyze a simplified, but instructive scenario. Our starting
point is the information Hamiltonian for all fields, Eq.\ \ref{eq:total-information-source},
which we marginalize with respect to the $\beta$ field,

\begin{eqnarray}
 &  & \mathcal{H}(d,\varphi,\chi,\overline{\chi})\nonumber \\
 & = & -\ln\int\mathcal{D}\beta\,e^{-\mathcal{H}(d,\psi)}\nonumber \\
 & = & -\ln\int\mathcal{D}\beta\,e^{-\{Q,\theta(\psi)\}-\mathcal{H}(d|\varphi)-\mathcal{H}(\varphi_{0})}\nonumber \\
 & = & -\ln\int\mathcal{D}\beta\,e^{i\beta^{\dagger}G[\varphi]-\frac{1}{2}\beta^{\dagger}\Xi\beta-\bar{\chi}^{\dagger}G'[\varphi]\,\chi-\mathcal{H}(d|\varphi)-\mathcal{H}(\varphi_{0})}\nonumber \\
 & \widehat{=} & -\ln e^{-\frac{1}{2}G[\varphi]^{\dagger}\Xi^{-1}G[\varphi]-\bar{\chi}^{\dagger}G'[\varphi]\,\chi-\mathcal{H}(d|\varphi)-\mathcal{H}(\varphi_{0})}\nonumber \\
 & = & \frac{1}{2}G[\varphi]^{\dagger}\Xi^{-1}G[\varphi]+\bar{\chi}^{\dagger}G'[\varphi]\,\chi+\mathcal{H}(d|\varphi)+\mathcal{H}(\varphi_{0}).\nonumber \\
\label{eq:H-without-beta}
\end{eqnarray}
The information Hamiltonian contains now, in this order, terms that
represent the excitation noise statistics $\mathcal{G}(\xi,\Xi)$
(as $\xi=G[\varphi]$), the functional determinant of the dynamics
(represented with help of fermionic fields), the measurement information
$\mathcal{H}(d|\varphi)$, and the information on the initial condition
$\mathcal{H}(\varphi_{0})$.

We assume the system $\varphi$ to be initially $\varphi(\cdot,0)=\varphi_{0}$
at $t=0$ and to obey Eq.\ \ref{eq:TimeDep.SDE} afterwards with
$\xi\hookleftarrow\mathcal{G}(\xi,\mathbb{1})$, \emph{i.e.} $\Xi=\mathbb{1}$.
We can then define a classical field $\varphi_{\text{cl}}$ that obeys
the excitation-free dynamics
\begin{equation}
\partial_{t}\varphi_{\text{cl}}(x)=F[\varphi_{\text{cl}}](x)
\end{equation}
and a deviation $\varepsilon:=\varphi-\varphi_{\text{cl}}$ from this,
which evolves according to
\begin{eqnarray}
\varepsilon(\cdot,0) & = & 0\text{ and}\\
\partial_{t}\varepsilon & = & F[\varphi_{\text{cl}}+\varepsilon]-F[\varphi_{\text{cl}}]+\xi\nonumber \\
 & = & \underbrace{\frac{\partial F[\varphi_{\text{cl}}]}{\partial\varphi_{\text{cl}}}}_{=:A}\,\varepsilon+\xi+\mathcal{O}(\varepsilon^{2}).
\end{eqnarray}
Here, we performed a first order expansion in the deviation field.
Furthermore, we assume that only a sufficiently short period after
$t=0$ is considered such that second order effects in $\varepsilon$
as well as any time dependence of $A$ can be ignored. For this period,
we have the solution
\begin{equation}
\varepsilon_{t}=\int_{0}^{t}d\tau\,e^{A(t-\tau)}\xi_{t'}.
\end{equation}

Further, we imagine that a system measurement at time $t=t_{\text{o}}$
probes perfectly a normalized eigendirection $b$ of $A$, i.e. that
we get noiseless data according to
\begin{equation}
d=R\varphi=b^{\hat{\dagger}}\varepsilon(\cdot,t_{\text{o}}).
\end{equation}
Here, $R_{1\,(\vec{x},t)}:=b_{\vec{x}}\delta(t-t_{\text{o}})$ is
the linear measurement operator, $b$ fulfills
\begin{eqnarray}
A\,b & = & \lambda_{b}\,b,
\end{eqnarray}
with $\lambda_{b}$ the corresponding eigenvalue, and $\hat{\dagger}$
denoting the adjoint with respect to spatial coordinates only. $\lambda_{b}$
is also the Lijapunov coefficient of the dynamical mode $b$, which
is stable for $\lambda_{b}<0$ and unstable for $\lambda_{b}>0$.
The latter is a prerequisite for chaos.

Finally, to exclude any further complications, we assume that $A$
can be fully expressed in terms of a set of such orthonormal eigenmodes,
\begin{equation}
A=\sum_{a}\lambda_{a}\,a\,a^{\hat{\dagger}}\text{ with }a^{\hat{\dagger}}a'=\delta_{aa'}.
\end{equation}

Now, we are in a convenient position to work out our knowledge on
$\varepsilon$ for all times for which our idealizing assumptions
hold.

\begin{figure*}
\includegraphics[width=0.333\textwidth]{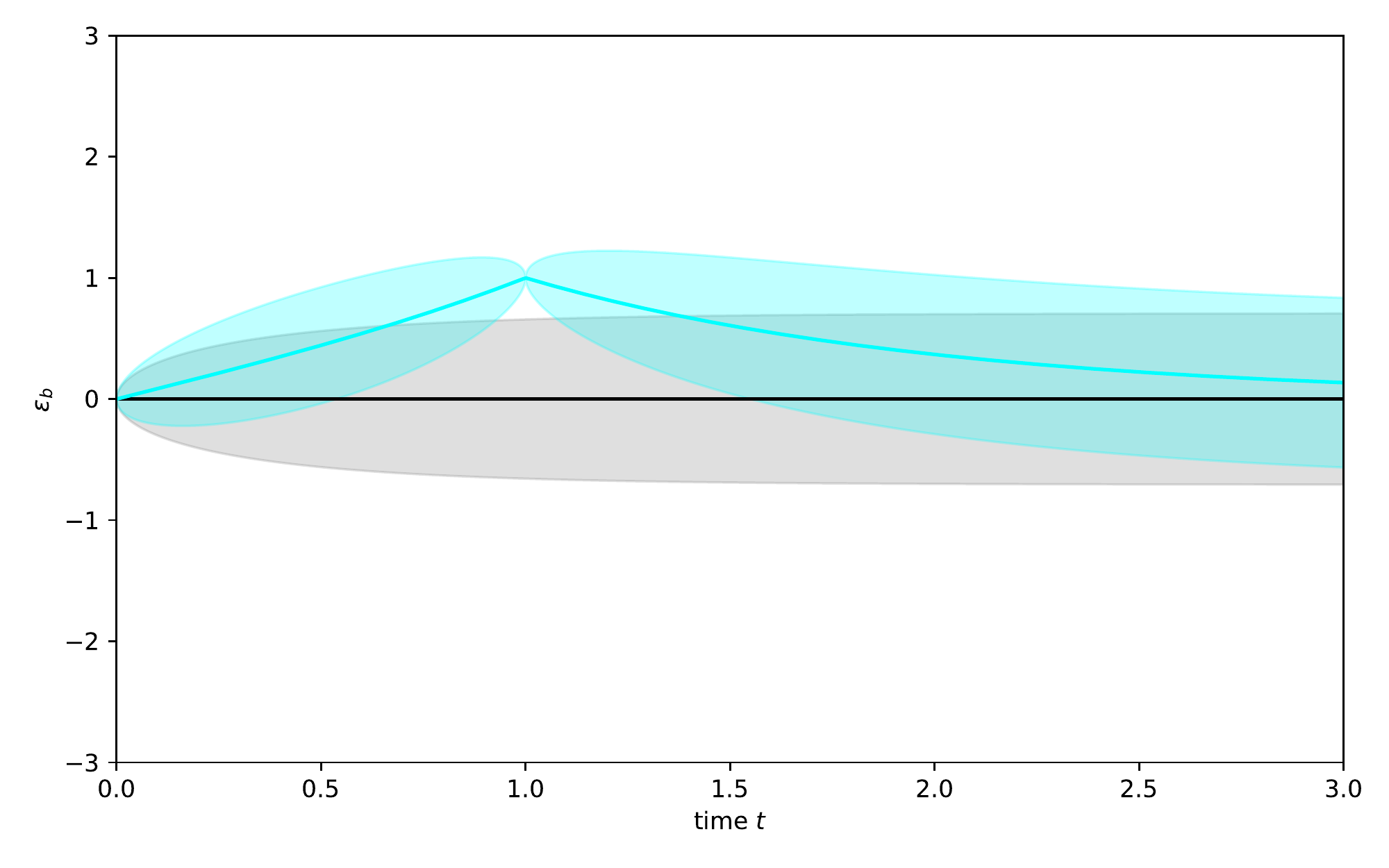}\includegraphics[width=0.333\textwidth]{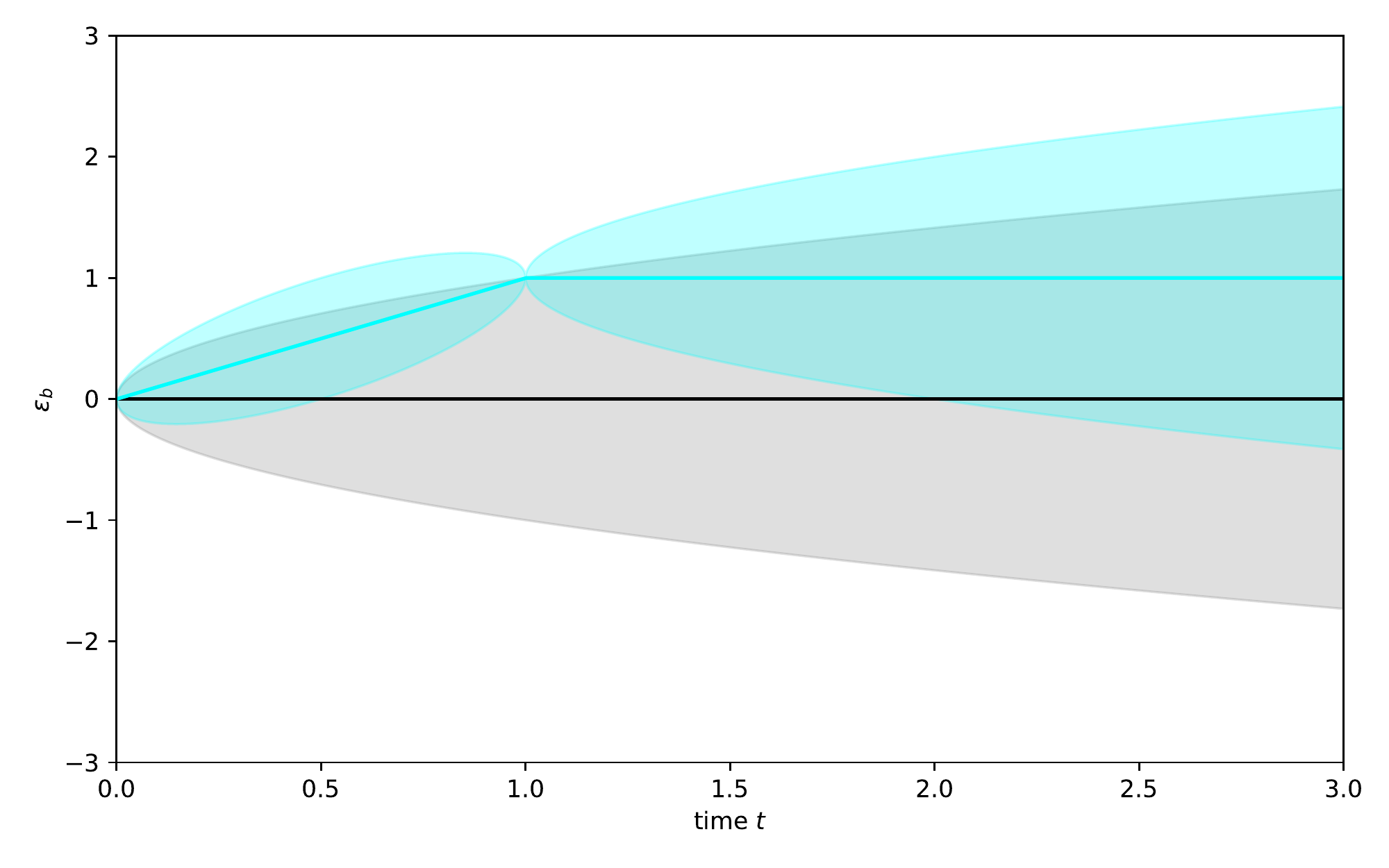}\includegraphics[width=0.333\textwidth]{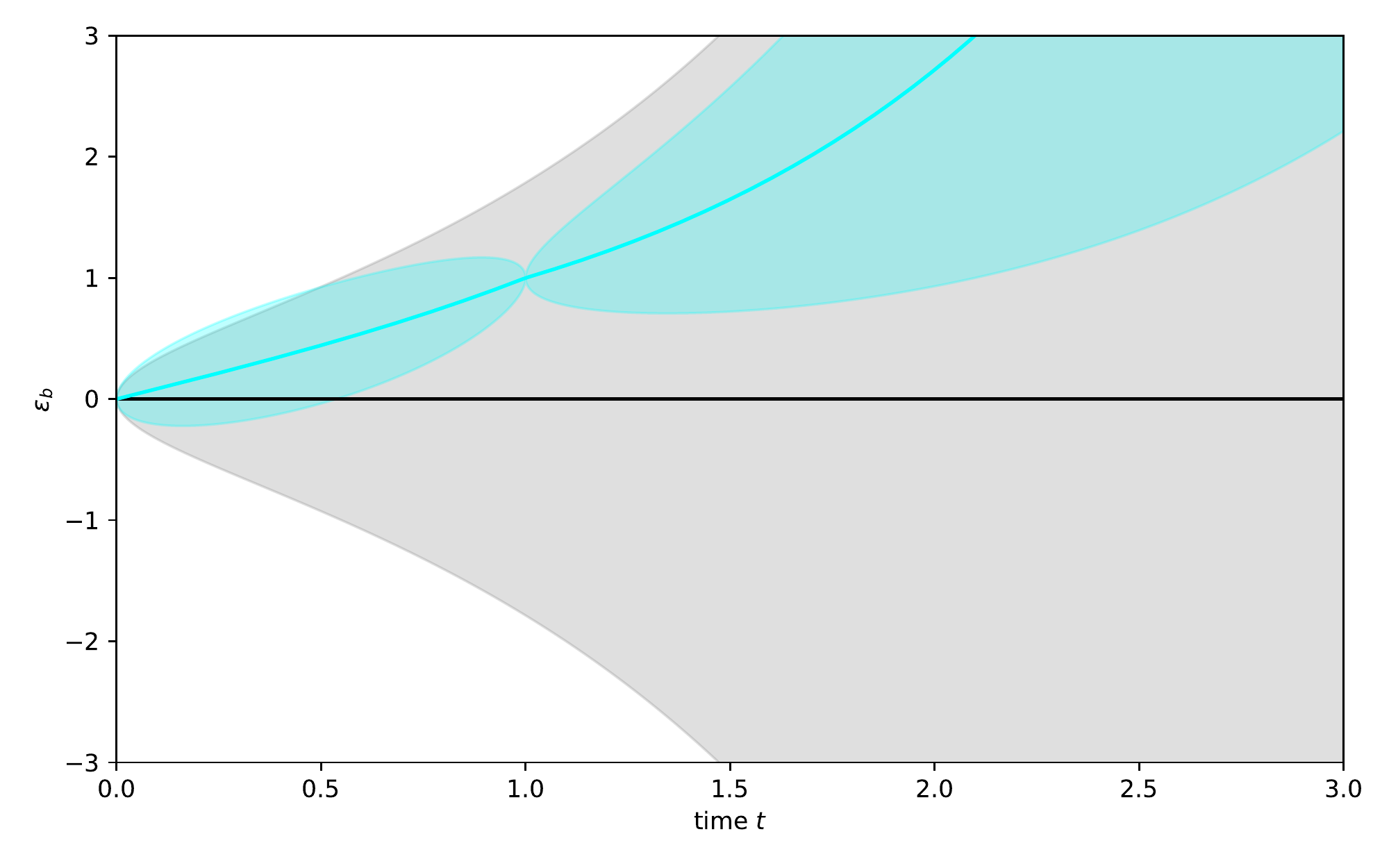}

\includegraphics[width=0.333\textwidth]{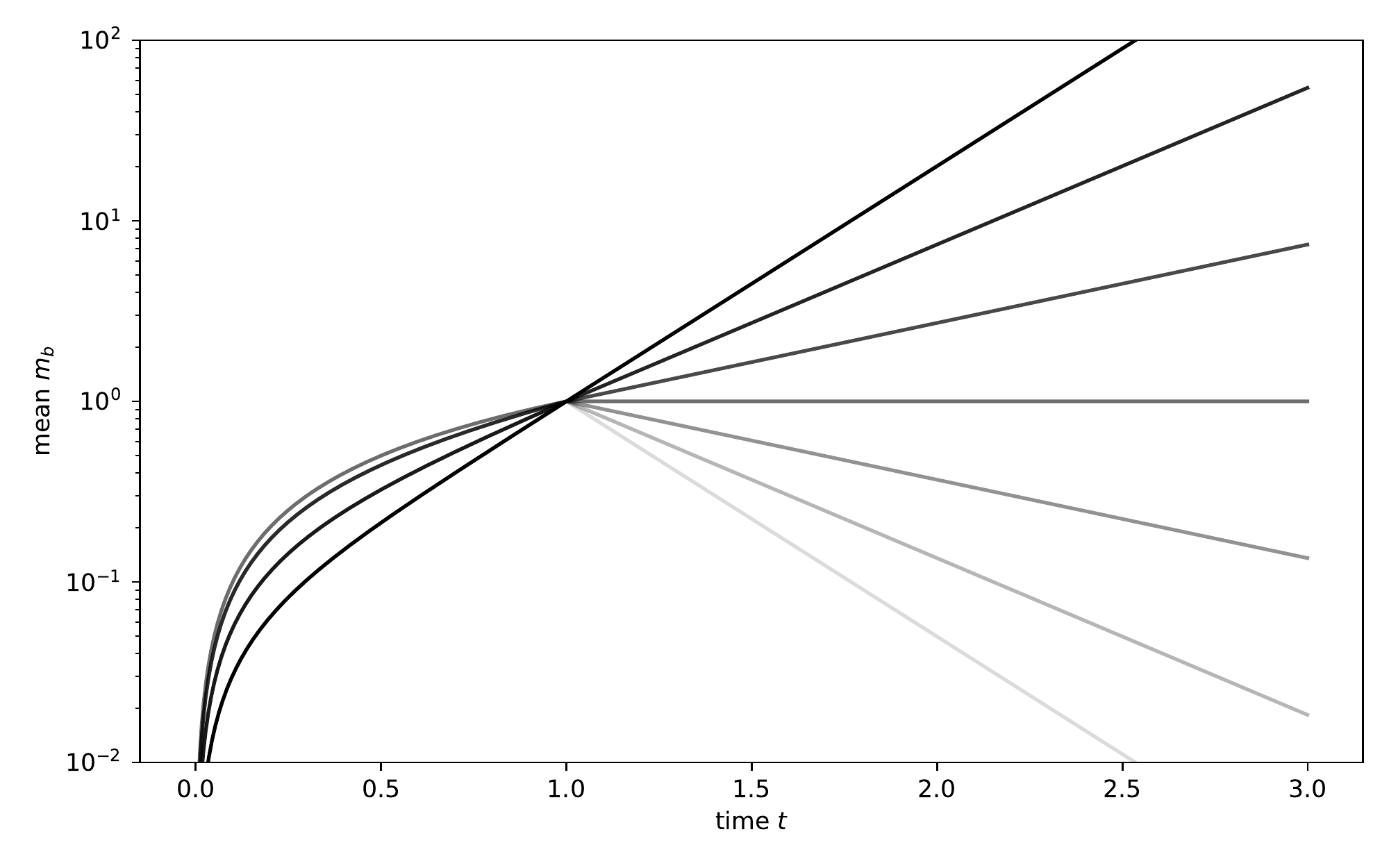}\includegraphics[width=0.333\textwidth]{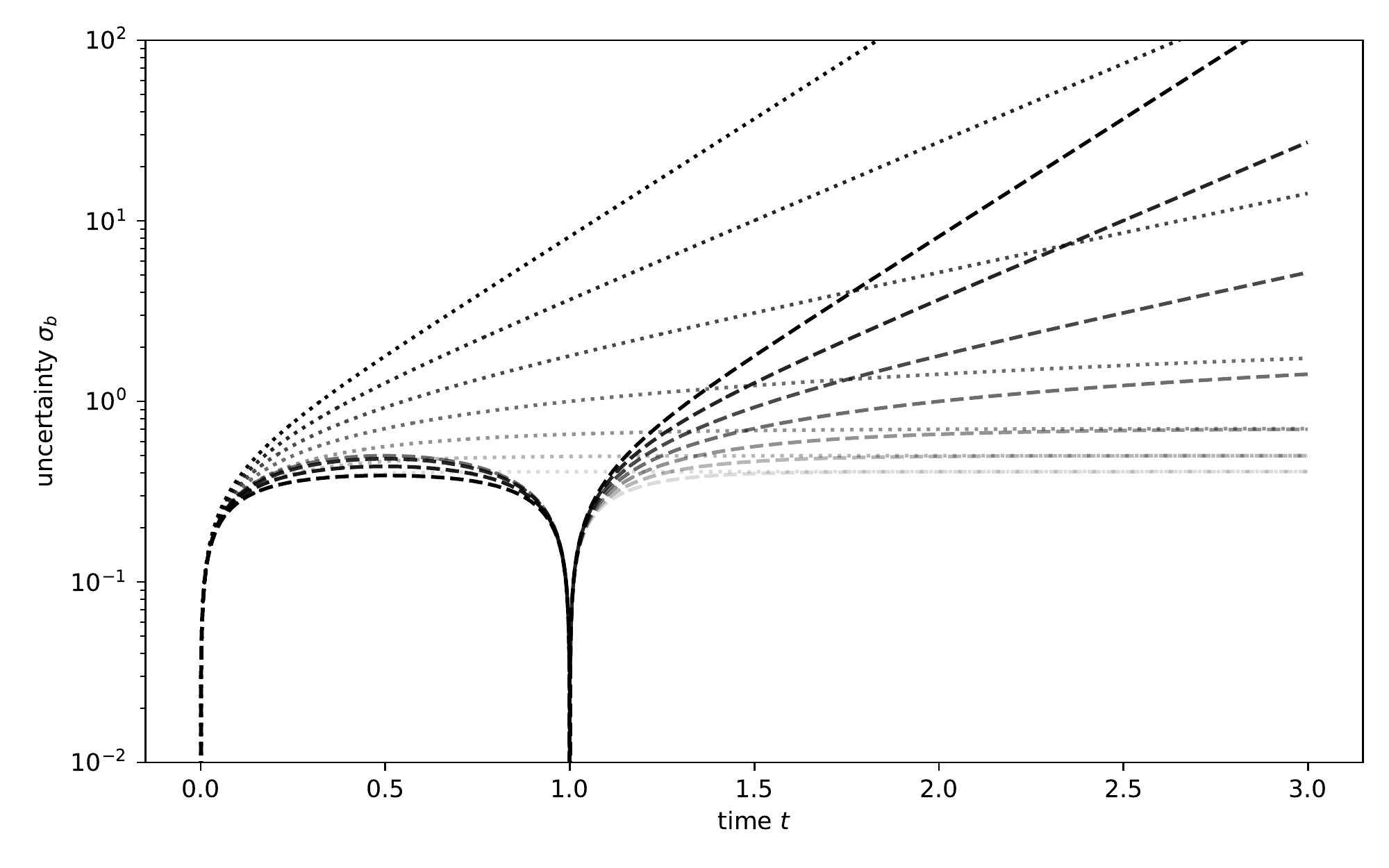}\includegraphics[width=0.333\textwidth]{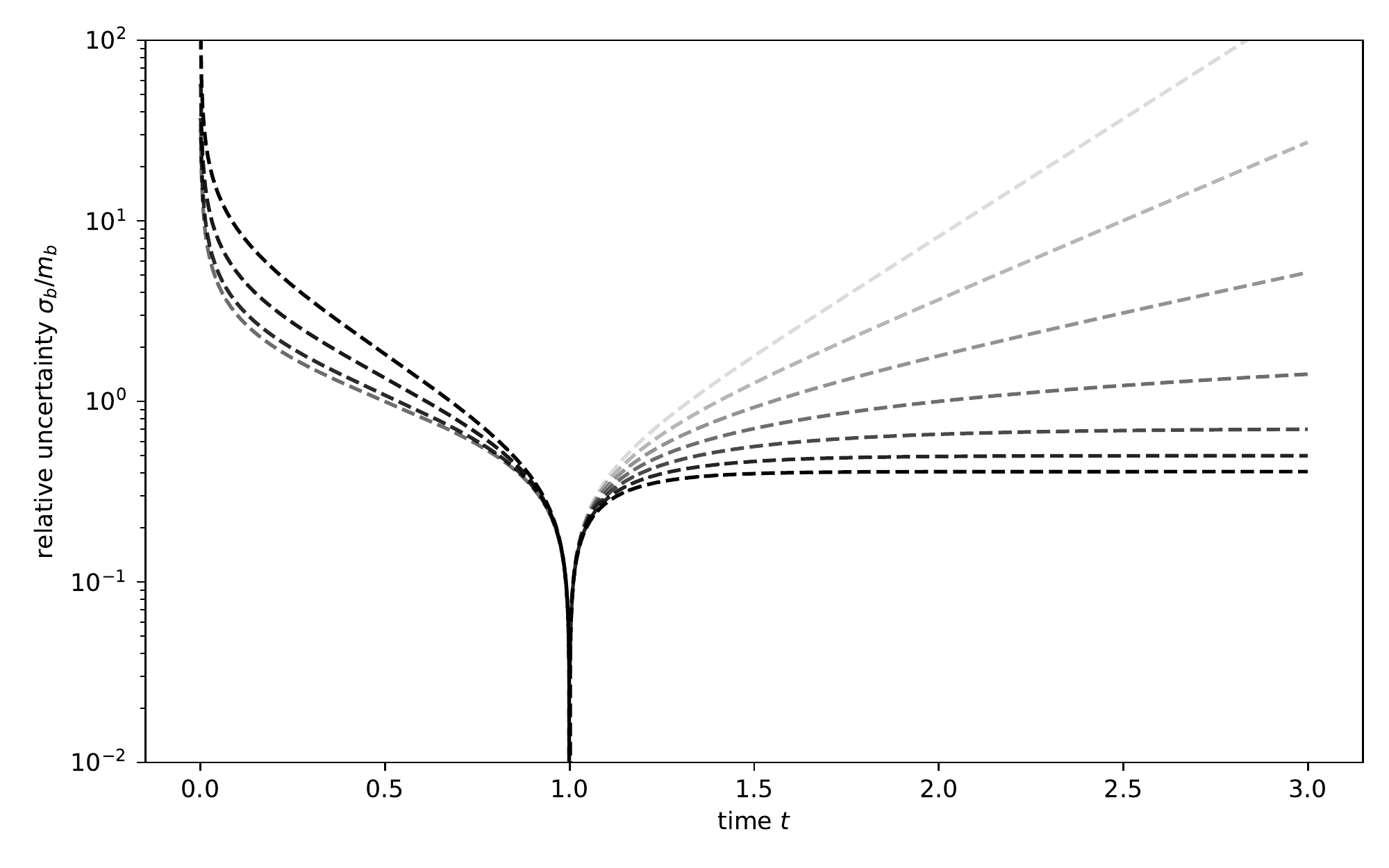}

\caption{Illustration of the knowledge on a measured system mode $b.$ Top
row: A priori (gray) and a posteriori (cyan) field mean (lines) and
one sigma uncertainty (shaded) for an Ornstein-Uhlenbeck process (left,
$\lambda_{b}=-1)$, a Wiener process (middle, $\lambda_{b}=0)$, and
a chaotic process (right, $\lambda_{b}=1$) of a system eigenmode
$b$ after one perfect measurement at $t_{o}=1$. Bottom row: The
same, but on logarithm scales and for Liapunov exponents $\lambda_{b}=-3$,
$-2$, $-1$, $0$, $1$, $2$, and $3$, as displayed in colors ranging
from light to dark gray in this order (\emph{i.e. }strongest chaos
is shown in black). Left: Posterior mean. Middle: Uncertainty of prior
(dotted) and posterior (dashed). Right: Relative posterior uncertainty.
\label{fig:Uncertainty-of-a}}
\end{figure*}

A priori, the deviation evolves with an average 
\begin{equation}
\overline{\varepsilon}_{t}:=\left\langle \varepsilon_{t}\right\rangle _{(\xi)}=\int_{0}^{t}d\tau\,e^{A(t-\tau)}\underbrace{\left\langle \xi_{\tau}\right\rangle _{(\xi)}}_{=0}=0
\end{equation}
and an dispersion, most conventiently expressed in the eigenbasis
of $A$, of 
\begin{eqnarray}
E_{(a,t)(a',t')} & := & \left\langle a^{\hat{\dagger}}\varepsilon_{t}\varepsilon_{t^{'}}^{\hat{\dagger}}a'\right\rangle _{(\xi)}\nonumber \\
 & = & \int_{0}^{t}d\tau\,\int_{0}^{t'}d\tau'\,a^{\hat{\dagger}}e^{A(t-\tau)}\times\nonumber \\
 &  & \underbrace{\left\langle \xi_{\tau}\xi_{\tau'}^{\hat{\dagger}}\right\rangle _{(\xi)}}_{=\delta(\tau-\tau')\,\mathbb{1}}e^{A^{\hat{\dagger}}(t'-\tau')}a'\nonumber \\
 & = & \int_{0}^{\text{min}(t,t')}d\tau\,\,e^{\lambda_{a}(t-\tau)}a^{\hat{\dagger}}a'\,e^{\lambda_{a'}(t'-\tau)}\nonumber \\
 & = & e^{\lambda_{a}(t+t')}\delta_{aa'}\left[1-e^{\lambda_{a}\text{min}(t,t')}\right]\,\left(2\lambda_{a}\right)^{-1}\nonumber \\
 & = & \delta_{aa'}\underbrace{\left[e^{\lambda_{a}(t+t')}-e^{\lambda_{a}|t-t'|}\right]\left(2\lambda_{a}\right)^{-1}}_{=:f_{a}(t,t')}\,.
\end{eqnarray}
We introduced here with $f_{a}(t,t'):=\langle a^{\hat{\dagger}}\varepsilon_{t}\varepsilon_{t^{'}}^{\hat{\dagger}}a\rangle_{(\xi)}$
the a priori temporal correlation function of a field eigenmode $a$.
Since both, the dynamics as well as the measurement, keep the eigenmodes
separate in our illustrative example, we only obtain additional information
on the mode $b$ from our measurement. This is given according to
Eq.\ \ref{eq:WFposterior} by the posterior
\begin{eqnarray}
\mathcal{P}(\varepsilon|d) & = & \mathcal{G}(\varepsilon-m,D)
\end{eqnarray}
with posterior mean
\begin{equation}
m=E\,R^{\dagger}\left(RE\,R^{\dagger}\right)^{-1}d
\end{equation}
 and posterior uncertainty
\begin{equation}
D=E-E\,R^{\dagger}\left(RE\,R^{\dagger}\right)^{-1}R\,E,
\end{equation}
which follow respectively from Eqs.\ \ref{eq:m-data-space} and \ref{eq:Dprop-data-space}
for the limit of vanishing noise covariance $N$. Expressing these
in the eigenbasis of $A$ gives
\begin{eqnarray}
m_{a}(t) & := & a^{\hat{\dagger}}m(\cdot,t)\nonumber \\
 & = & \delta_{ab}\frac{f_{b}(t,t_{\text{o}})}{f_{b}(t_{\text{o}},t_{\text{o}})}\,d
\end{eqnarray}
and 
\begin{eqnarray}
D_{(a,t)(a',t')} & = & \delta_{aa'}\left[f_{a}(t,t')-\delta_{ab}\frac{f_{b}(t,t_{\text{o}})f_{b}(t',t_{\text{o}})}{f_{b}(t_{\text{o}},t_{\text{o}})}\right].\nonumber \\
\end{eqnarray}
Fig.\ \ref{fig:Uncertainty-of-a} shows the mean and uncertainty
dispersion of the measured mode for various values of $\lambda_{b}$.
The correlation between different modes $a\neq a'$ vanishes and therefore
any mode $a\neq b$ behaves like a prior mode shown in grey in Fig.\ \ref{fig:Uncertainty-of-a}.
For the measured mode $b$, the propagator is in general non-zero,
but vanishes for times separated by the observation, e.g. $D_{(b,t)(b,t')}=0$
for $t<t_{\text{o}}<t'$, as one can easily verify:
\begin{eqnarray}
 &  & D_{(b,t)(b,t')}\times f_{b}(t_{\text{o}},t_{\text{o}})\nonumber \\
 & = & f_{b}(t,t')\,f_{b}(t_{\text{o}},t_{\text{o}})-f_{b}(t,t_{\text{o}})f_{b}(t',t_{\text{o}})\nonumber \\
 & = & \left[e^{\lambda_{b}(t+t')}-e^{\lambda_{b}|t-t'|}\right]\,\left[e^{2\lambda_{b}t_{\text{o}}}-1\right]-\nonumber \\
 &  & \left[e^{\lambda_{b}(t+t_{\text{o}})}-e^{\lambda_{b}|t-t_{\text{o}}|}\right]\,\left[e^{\lambda_{b}(t_{\text{o}}+t')}-e^{\lambda_{b}|t_{\text{o}}-t'|}\right]\nonumber \\
 & = & \left[e^{\lambda_{b}(t+t')}-e^{\lambda_{b}(t'-t)}\right]\,\left[e^{2\lambda_{b}t_{\text{o}}}-1\right]-\nonumber \\
 &  & \left[e^{\lambda_{b}(t+t_{\text{o}})}-e^{\lambda_{b}(t_{\text{o}}-t)}\right]\,\left[e^{\lambda_{b}(t_{\text{o}}+t')}-e^{\lambda_{b}(t'-t_{\text{o}})}\right]\nonumber \\
 & = & \left[e^{\lambda_{b}(t+t'+2t_{\text{o}})}-e^{\lambda_{b}(t+t')}-e^{\lambda_{b}(t'-t+2t_{\text{o}})}+e^{\lambda_{b}(t'-t)}\right]-\nonumber \\
 &  & \left[e^{\lambda_{b}(t+t'+2t_{\text{o}})}-e^{\lambda_{b}(t+t')}-e^{\lambda_{b}(2t_{\text{o}}+t'-t)}+e^{\lambda_{b}(t'-t)}\right]\nonumber \\
 & = & 0\label{eq:D_vanishes}
\end{eqnarray}
Thus, the perfect measurement introduces a so-called Markov-blanket,
which separates the periods before and after it from each other. Knowing
anything above earlier times than $t_{\text{o}}$ does not inform
about later times, as the measuremnt at $t_{\text{o}}$ provides the
only relevant constraint for the later period. The equal time uncertainty
of the measured mode is
\begin{eqnarray}
D_{(b,t)(b,t)} & = & f_{b}(t,t)-\frac{f_{b}(t,t_{\text{o}})f_{b}(t,t_{\text{o}})}{f_{b}(t_{\text{o}},t_{\text{o}})}\nonumber \\
 & = & \frac{1}{2\lambda_{b}}\left[e^{2\lambda_{b}t}-1-\frac{\left[e^{\lambda_{b}(t+t_{\text{o}})}-e^{\lambda_{b}|t-t_{\text{o}}|}\right]^{2}}{2\lambda_{b}\left[e^{2\lambda_{b}t_{\text{o}}}-1\right]}\right].\nonumber \\
\end{eqnarray}
Fig.\ \ref{fig:Uncertainty-of-a} shows this for a number of instructive
values of $\lambda_{b}.$ The impact of the Liapunov exponent on the
predictability of the system is clearly visible. As larger the Liapunov
exponent, as faster grow uncertainties. This can be seen by comparison
of the top panels or by inspection of the bottom middle panel of Fig.\ \ref{fig:Uncertainty-of-a}.
Thus, chaos, which implies the existence of positive Liapunov exponents,
makes field inference more difficult. This, however, is only true
on an absolute scale. If one considers relative uncertainties, as
also displayed in Fig.\ \ref{fig:Uncertainty-of-a} on the bottom
right, then it turns out that these grow slowest for the more unstable
modes. This is the memory effect of chaotic systems, which can remember
small initial disturbances for long, if not infinite times.

To simplify the system further, we concentrate first on the case $\lambda_{b}=0$,
which corresponds to a Wiener process. For this we get
\begin{eqnarray}
f_{a}(t,t') & = & \text{min}(t,t'),
\end{eqnarray}
implying a posterior mean of 
\begin{eqnarray}
m_{a}(t) & = & \delta_{ab}\text{min}(\nicefrac{t}{t_{\text{o}}},1)\,d
\end{eqnarray}
and an information propagator of
\begin{eqnarray}
D_{(a,t)(a',t')} & = & \delta_{aa'}\left[\min(t,t')-\delta_{ab}\frac{\min(t,t_{\text{o}})\min(t',t_{\text{o}})}{t_{\text{o}}}\right].\nonumber \\
\end{eqnarray}
This provides the equal time uncertainty for our measured mode $b$
\begin{eqnarray}
D_{(b,t)(b,t)} & = & t-\frac{\min(t,t_{\text{o}})^{2}}{t_{\text{o}}}=\begin{cases}
t\,(1-\nicefrac{t}{t_{\text{o}})} & t<t_{\text{o}}\\
t-t_{\text{o}} & t\ge t_{\text{o}}
\end{cases},\;\;\;\;\;\;\;\;\;\;
\end{eqnarray}
which is also shown in Fig.\ \ref{fig:Uncertainty-of-a} in both
middle panels. This scenario with $\lambda_{b}=0$ corresponds to
a Wiener process, which sits on the boundary between the stable Ornstein-Uhlenbeck
process with $\lambda_{b}<0$ and the instability of chaos with $\lambda_{b}>0$.
This marginal stable case should now be taken into the non-linear
regime.

\subsection{Idealized Non-Linear Dynamics\label{subsec:Idealized-Non-Linear-Dynamics}}

We saw that the posterior uncertainty is a good indicator for the
difficulty to predict the field at locations or times where or when
it was not measured. This holds -- modulo some corrections -- also
in the case of non-linear dynamics, which introduces non-Gaussianities
into the field statistics.

In order to investigate such a non-Gaussian example, we extend the
previous case with $\lambda_{b}=0$ to the next order in $\varepsilon$,
while still assuming that all modes are dynamically decoupled (up
to that order), such that we only need to concentrate on the dynamics
of $\varepsilon_{b}(t):=b^{\hat{\dagger}}\varepsilon(\cdot,t),$
\begin{equation}
\partial_{t}\varepsilon_{b}=\frac{1}{2}\,\mu_{b}\,\varepsilon_{b}^{2}+\xi+\mathcal{O}(\varepsilon_{b}^{3}),
\end{equation}
where again $\hat{\dagger}$ denotes an integration in position space
only. This mode will exhibit an infinite posterior mean for times
larger than $t_{\text{o}}$. To understand why, let us first investigate
the noise free solution of $\partial_{t}\varepsilon_{b}=\frac{1}{2}\,\mu_{b}\,\varepsilon_{b}^{2}$
for some finite starting value $\varepsilon(t_{\text{i}})=\varepsilon_{\text{i}}$
at $t_{\text{i}}>t_{\text{o}}$. This might have been created by an
excitation fluctuation during the period $[t_{\text{o}},t_{\text{i}}]$
for which always a potentially tiny, but finite probability exists.
The free solution after $t_{\text{i }}$is given by
\begin{equation}
\varepsilon_{b}(t)=\frac{\varepsilon_{\text{i}}}{1-\frac{1}{2}\,\varepsilon_{\text{i}}\,\mu_{b}(t-t_{\text{i}})},\label{eq:explosion}
\end{equation}
which develops a singularity for $\varepsilon_{\text{i}}\,\mu_{b}>0$
in the finite period $\tau=2/(\varepsilon_{\text{i}}\,\mu_{b})$.
Thus, there is a finite probability that at time $t_{\text{s}}=t_{\text{i}}+\tau$
the system is at infinity, and this lets also the expectation value
of $\varepsilon$ diverge for $t_{\text{s}}$. This moment, when the
expectation value has diverged, can be made arbitrarily close to $t_{\text{o}}$,
as the Gaussian fluctuations in $\xi$ permit to reach any necessary
$\varepsilon_{\text{i}}$ at say $t_{\text{i}}=(t_{\text{s}}-t_{\text{o}})/2=\tau$
with a small, but finite probability, where $\varepsilon_{\text{i}}=2/(\tau\,\mu_{b})=4/[(t_{\text{s}}-t_{\text{o}})\,\mu_{b}]$.

For times $t\in[0,t_{\text{o}}]$, in between the moments when the
two data points were measured, the posterior mean should stay finite.
The reason is that any a priori possible trajectory diverging to (plus)
infinity (for $\mu_{b}>0$) during this period is excluded a posteriori
by the data point $(t,\varepsilon_{b})=(1,1)$. Such trajectories
could not have taken place, as the dynamics does not permit trajectories
to return from (positive) infinite values to finite ones, since that
would require an infinite large (negative) excitation, which does
have a probability of zero.

Let us assume that for the period $t\in[0,t_{\text{o}}]$ the second
order approximation of the dynamical equation holds. We then have
\begin{eqnarray}
G[\varepsilon_{b}] & = & \partial_{t}\varepsilon_{b}-\frac{1}{2}\,\mu_{b}\,\varepsilon_{b}^{2},
\end{eqnarray}
and therefore 
\begin{eqnarray}
\frac{\delta G[\varepsilon_{b}]}{\delta\varepsilon_{b}} & = & \partial_{t}-\mu_{b}\,\varepsilon_{b}.
\end{eqnarray}
Inserting this into Eq.\ \ref{eq:H-without-beta} yields
\begin{eqnarray}
\mathcal{H}(d,\varphi,\chi,\overline{\chi}) & = & \frac{1}{2}\left[\partial_{t}\varepsilon_{b}-\frac{1}{2}\,\mu_{b}\,\varepsilon_{b}^{2}\right]^{\dagger}\left[\partial_{t}\varepsilon_{b}-\frac{1}{2}\,\mu_{b}\,\varepsilon_{b}^{2}\right]\nonumber \\
 &  & +\bar{\chi}^{\dagger}\left[\partial_{t}-\mu_{b}\,\varepsilon_{b}\right]\,\chi+\mathcal{H}(d|\varphi)+\mathcal{H}(\varphi_{0}).\nonumber \\
 & = & \mathcal{H_{\text{free}}}(d,\varphi,\chi,\overline{\chi})+\mathcal{H_{\text{int}}}(d,\varphi,\chi,\overline{\chi})
\end{eqnarray}
with
\begin{eqnarray}
\mathcal{H_{\text{free}}}(d,\varphi,\chi,\overline{\chi}) & = & \frac{1}{2}\varepsilon_{b}^{\dagger}\partial_{t}^{\dagger}\partial_{t}\varepsilon_{b}+\bar{\chi}^{\dagger}\partial_{t}\,\chi+\mathcal{H}(d|\varphi)+\mathcal{H}(\varphi_{0})\nonumber \\
\mathcal{H_{\text{int}}}(d,\varphi,\chi,\overline{\chi}) & = & -\mu_{b}\left[\varepsilon_{b}^{2}\right]^{\dagger}\partial_{t}\,\varepsilon_{b}+\frac{\mu_{b}^{2}}{8}\left[\varepsilon_{b}^{2}\right]^{\dagger}\left[\varepsilon_{b}^{2}\right]\nonumber \\
 &  & -\mu_{b}\bar{\chi}^{\dagger}\left(\varepsilon_{b}\,\chi\right).
\end{eqnarray}
The free information Hamiltonian $\mathcal{H_{\text{free}}}(d,\varphi,\chi,\overline{\chi})$
defines the Wiener process field inference problem we addressed before,
and has the classical field as well as the bosonic and fermionic propagators
given by 
\begin{eqnarray}
m(t) & = & \text{min}(\nicefrac{t}{t_{\text{o}}},1)\,d=\frac{t\,d}{t_{\text{o}}}\text{ for }t<t_{\text{o}}\nonumber \\
 & = & \WF\\
D_{tt'}^{\text{b}} & = & \min(t,t')-\frac{\min(t,t_{\text{o}})\min(t',t_{\text{o}})}{t_{\text{o}}}\nonumber \\
 & = & \min(t,t')-\frac{t\,t'}{t_{\text{o}}}\text{ for }t,t'<t_{\text{o}}\nonumber \\
 & = & \propagator\\
D_{tt'}^{\text{f}} & = & \left[\delta(t-t')\partial_{t'}\right]^{-1}=\theta(t-t')\nonumber \\
 & = & \ghost
\end{eqnarray}
respectively. Here, we introduced their Feynman diagram representation
as well. The Fermionic propagator is the inverse of $\delta(t-t')\partial_{t'}$
as is verified by
\begin{eqnarray}
\int\text{d}t'\,\left[\delta(t-t')\partial_{t'}\right]D_{t't''}^{\text{f}} & = & \int\text{d}t'\,\left[\delta(t-t')\partial_{t'}\right]\theta(t'-t'')\nonumber \\
 & = & \int\text{d}t'\,\delta(t-t')\delta(t'-t'')\nonumber \\
 & = & \delta(t-t'')=\mathbb{1}_{tt''}.
\end{eqnarray}
 
\begin{figure*}
\includegraphics[width=0.333\textwidth]{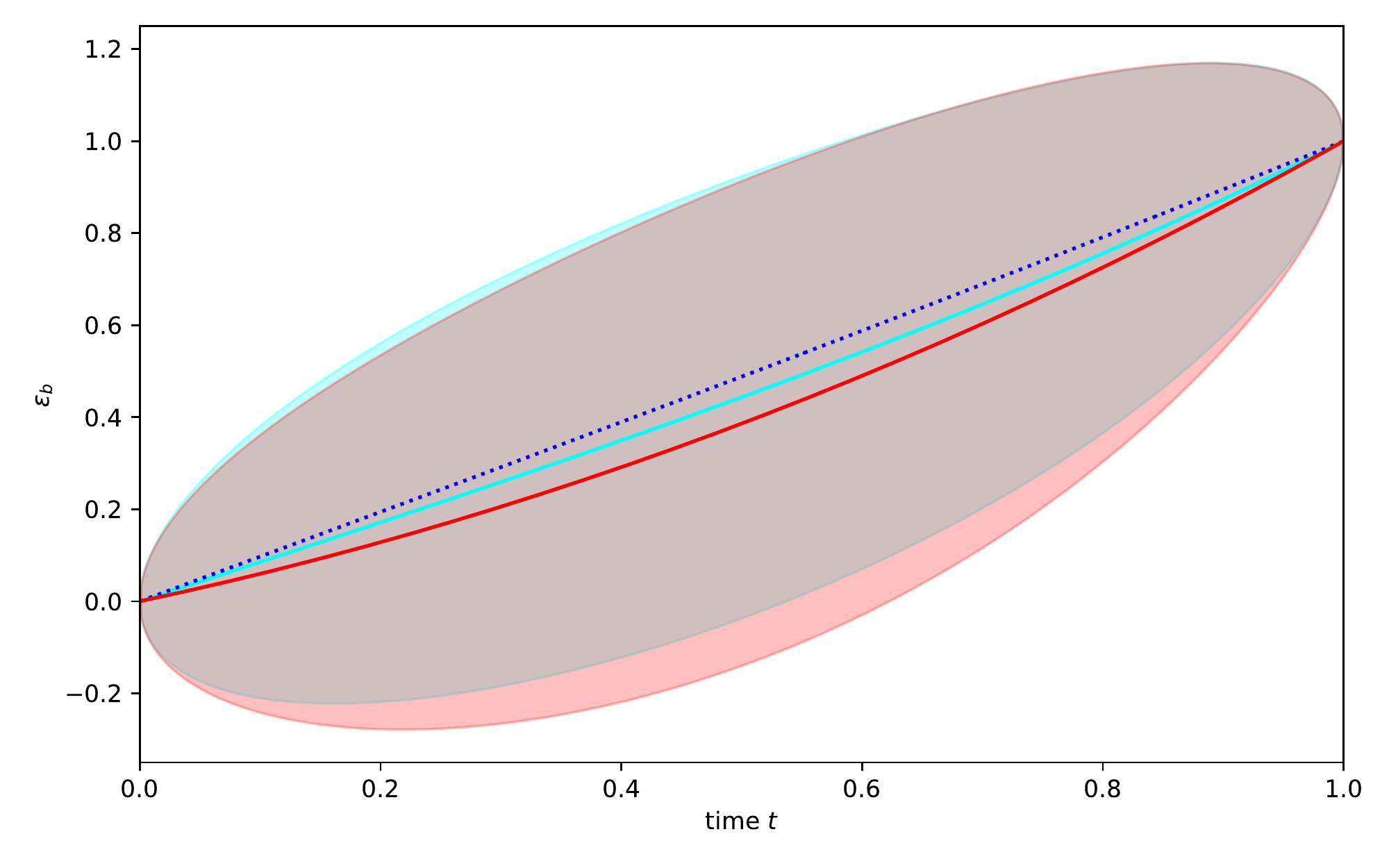}\includegraphics[width=0.333\textwidth]{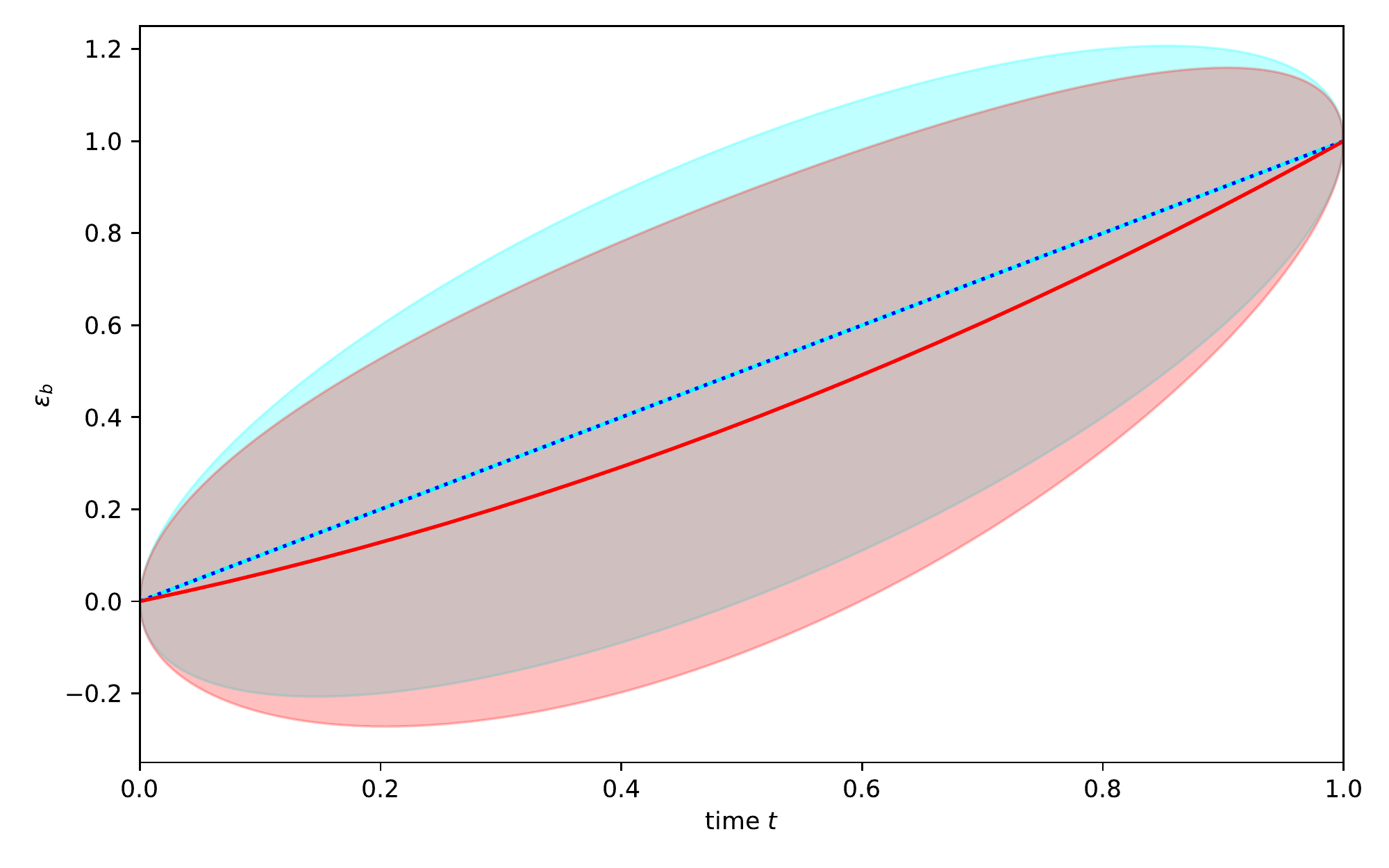}\includegraphics[width=0.333\textwidth]{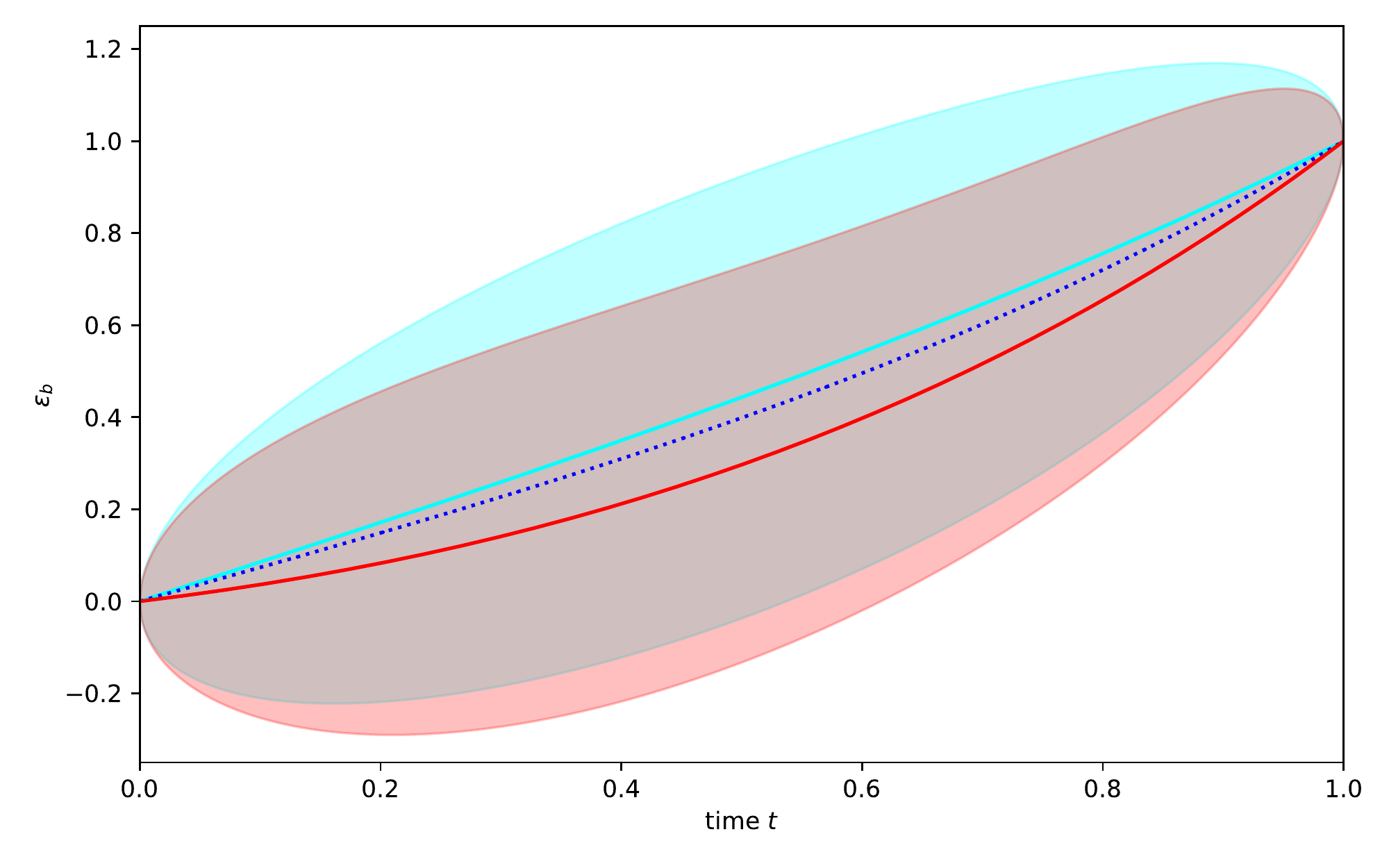}

\caption{Like top row of Fig.\ \ref{fig:Uncertainty-of-a} just for the non-linear
system defined by Eq.\ \ref{eq:non-linear-system-G} within the period
$t\in[0,1]$ with first order bosonic and fermionic perturbation corrections
for $\mu_{b}=0.3$ in red, as in Fig.\ \ref{fig:Uncertainty-of-a}
without such non-linear corrections in cyan, and with only bosonic
corrections in blue (dotted, displayed without uncertainty). The three
panels display the cases $\lambda_{b}=-1$ (left), $\lambda_{b}=0$
(middle), and $\lambda_{b}=1$ (right). Note that the a priori mean
and uncertainty dispersion are both infinite for any time $t>0$,
as without the measurement, trajectories reaching positive infinity
within finite times are not excluded from the ensemble of permitted
possibilities. \label{fig:Uncertainty-of-a-1}}
\end{figure*}

The interacting Hamiltonian $\mathcal{H_{\text{int}}}(d,\varphi,\chi,\overline{\chi})$
provides the following interaction vertices
\begin{eqnarray}
\diaA & = & -2\,\mu_{b}\delta(t_{1}-t_{2})\,\delta(t_{2}-t_{3})\,\left[\partial_{t_{1}}+\partial_{t_{2}}+\partial_{t_{3}}\right]\nonumber \\
\\
\diaB & = & -3!\,\mu_{b}\delta(t_{1}-t_{2})\,\delta(t_{2}-t_{3})\\
\diaC & = & 3\mu_{b}^{2}\delta(t_{1}-t_{2})\,\delta(t_{2}-t_{3})\,\delta(t_{3}-t_{4}).
\end{eqnarray}
The integration over the time axis in Feynman diagrams can be restricted
to the interval $[0,t_{\text{o}}]$ as the propagator vanishes for
(exactly) one of the times being larger than $t_{\text{o}}$, see
Eq.\ \ref{eq:D_vanishes}.

To first order in $\mu_{b}$, the posterior mean and uncertainty dispersion
for $0\le t,t'\le t_{\text{o}}$ are then given by the Feynman diagrams
\begin{eqnarray}
\left\langle \varepsilon_{b}\right\rangle _{(\varepsilon_{b}|d)} & = & \WFc+\diaD+\nonumber \\
 &  & \diaE+\diaF+\mathcal{O}(\mu_{b}^{2})\nonumber \\
 & = & \frac{t}{t_{\text{o}}}\,d-\frac{3}{2}\,\mu_{b}t\,(t_{\text{o}}-t)+\mathcal{O}(\mu_{b}^{2})\\
\left\langle \varepsilon_{b}\varepsilon_{b}^{\dagger}\right\rangle _{(\varepsilon_{b}|d)}^{\text{c}} & = & \propagatorc+\diaG+\mathcal{O}(\mu_{b}^{2})\nonumber \\
 & = & \min(t,t')-\frac{t\,t'}{t_{\text{o}}}+\mathcal{O}(\mu_{b}^{2}),
\end{eqnarray}
see Appendix \ref{sec:Feynman-diagrams}. It turns out that all first
order diagrams (in $\mu_{b})$ with a bosonic three-vertex are zero.
The reason for this lies in the fact that these are all of a similar
form,
\begin{eqnarray}
 &  & \int_{0}^{t_{\text{o}}}\text{d}t_{1}\int_{0}^{t_{\text{o}}}\text{d}t_{2}\int_{0}^{t_{\text{o}}}\text{d}t_{3}\delta(t_{1}-t_{2})\,\delta(t_{2}-t_{3})\times\nonumber \\
 &  & \left[D_{tt_{3}}^{\text{b}}\partial_{t_{1}}g(t_{1},t_{2})+D_{tt_{3}}^{\text{b}}\partial_{t_{2}}g(t_{1},t_{2})+g(t_{1},t_{2})\partial_{t_{3}}D_{tt_{3}}^{\text{b}}\right]\nonumber \\
 & = & \int_{0}^{t_{\text{o}}}\text{d}t_{1}\left[D_{tt_{1}}^{\text{b}}\partial_{t_{1}}g(t_{1},t_{1})+g(t_{1},t_{1})\partial_{t_{1}}D_{tt_{1}}^{\text{b}}\right]\nonumber \\
 & = & \int_{0}^{t_{\text{o}}}\text{d}t_{1}D_{tt_{1}}^{\text{b}}\partial_{t_{1}}g(t_{1},t_{1})\nonumber \\
 &  & +\left[g(t_{1},t_{1})D_{tt_{1}}^{\text{b}}\right]_{t_{1}=0}^{t_{\text{o}}}-\int_{0}^{t_{\text{o}}}\text{d}t_{1}D_{tt_{1}}^{\text{b}}\partial_{t_{1}}g(t_{1},t_{1})\nonumber \\
 & = & g(t_{\text{o}},t_{\text{o}})D_{t_{\text{o}}t_{\text{o}}}^{\text{b}}-g(0,0)D_{00}^{\text{b}}=0,\label{eq:vansishing}
\end{eqnarray}
with $g(t_{1},t_{2})=\mu_{b}m_{t_{1}}m_{t_{2}}\text{, }\frac{1}{2}\mu_{b}D_{t_{1}t_{2}}^{\text{b}}$,
and $\mu_{b}m_{t_{1}}D_{t_{2}t'}^{\text{b}}$ respectively. All these
diagrams vanish, because $D_{t_{\text{o}}t_{\text{o}}}^{\text{b}}=D_{00}^{\text{b}}=0$.
Thus, to first order in $\mu_{b}$ only a correction due to the Fermionic
loop is necessary. This is negative (for positive $\mu_{b}$) as from
the sum over trajectories, which go through the initial data $(t_{\text{i}},\varepsilon_{b\text{i}})=(0,0)$
as well as through the later observed data $(t_{\text{o}},\varepsilon_{b\text{o}})=(1,1)$,
all the trajectories that diverge prematurely (within $t\in[0,t_{\text{o}}]$)
are excluded.

The posterior mean and uncertainty of the scenario with $\lambda_{b}=0$
and $\mu_{b}=0.3$ is displayed for $t\in[0,t_{\text{o}}]$ in the
middle panel of Fig.\ \ref{fig:Uncertainty-of-a-1} in red in comparison
to those for $\lambda_{b}=0$ and $\mu_{b}=0$ in cyan. It can there
be observed that the exclusion of the diverging trajectories by the
observation has made the ensemble of remaining trajectories staying
away from high values, which more easily diverge. Furthermore, this
effect is solely represented by the fermionic Feynman diagram, as
all bosonic corrections vanish (for $\lambda_{b}=0$) up to the considered
linear order in $\mu_{b}$. Thus, taking the functional determinant
into account, for which the fermionic fields were introduced, is important
in order to arrive at the correct posterior statistics. This effect
naturally arises in the here used Stratonovich formalism of stochastic
systems, and is less obvious in Îto's formalism.

Now, we are in a position to also work out the corrections in case
$\lambda\neq0$. In this case, we have
\begin{eqnarray}
G[\varepsilon_{b}] & = & \partial_{t}\varepsilon_{b}-\lambda_{b}\,\varepsilon_{b}-\frac{1}{2}\,\mu_{b}\,\varepsilon_{b}^{2}\label{eq:non-linear-system-G}
\end{eqnarray}
and
\begin{eqnarray}
\frac{\delta G[\varepsilon_{b}]}{\delta\varepsilon_{b}} & = & \partial_{t}-\lambda_{b}-\mu_{b}\,\varepsilon_{b}
\end{eqnarray}
such that now
\begin{eqnarray}
\mathcal{H_{\text{free}}}(d,\varphi,\chi,\overline{\chi}) & = & \frac{1}{2}\varepsilon_{b}^{\dagger}\left[\left(\partial_{t}-\lambda_{b}\right)^{\dagger}\left(\partial_{t}-\lambda_{b}\right)\right]\varepsilon_{b}\\
 &  & +\bar{\chi}^{\dagger}\left(\partial_{t}-\lambda_{b}\right)\,\chi+\mathcal{H}(d|\varphi)+\mathcal{H}(\varphi_{0})\nonumber \\
\mathcal{H_{\text{int}}}(d,\varphi,\chi,\overline{\chi}) & = & -\mu_{b}\left[\varepsilon_{b}^{2}\right]^{\dagger}\left(\partial_{t}-\lambda_{b}\right)\,\varepsilon_{b}+\frac{\mu_{b}^{2}}{8}\left[\varepsilon_{b}^{2}\right]^{\dagger}\left[\varepsilon_{b}^{2}\right]\nonumber \\
 &  & -\mu_{b}\bar{\chi}^{\dagger}\left(\varepsilon_{b}\,\chi\right)
\end{eqnarray}
and
\begin{eqnarray}
m(t) & = & \frac{f_{b}(t,t_{\text{o}})}{f_{b}(t_{\text{o}},t_{\text{o}})}\,d\text{ for }t<t_{\text{o}}\nonumber \\
 & = & \WF\\
D_{tt'}^{\text{b}} & = & f_{b}(t,t')-\frac{f_{b}(t,t_{\text{o}})f_{b}(t',t_{\text{o}})}{f_{b}(t_{\text{o}},t_{\text{o}})}\nonumber \\
 & = & \propagator\\
D_{tt'}^{\text{f}} & = & \left[\delta(t-t')\left(\partial_{t}-\lambda_{b}\right)\right]^{-1}\nonumber \\
 & = & \theta(t-t')e^{\lambda_{b}(t-t')}\nonumber \\
 & = & \ghost
\end{eqnarray}

where
\begin{eqnarray}
f_{b}(t,t') & = & \frac{e^{\lambda_{b}(t+t')}-e^{\lambda_{b}|t-t'|}}{2\lambda_{b}}.
\end{eqnarray}
 The Fermionic propagator for $\lambda\neq0$ is easily verified:

\begin{eqnarray}
 &  & \int\text{d}t'\delta(t-t')\left(\partial_{t'}-\lambda_{b}\right)D_{tt'}^{\text{f}}\nonumber \\
 & = & \int\text{d}t'\delta(t-t')\left(\partial_{t'}-\lambda_{b}\right)\left[\theta(t'-t'')e^{\lambda_{b}(t'-t'')}\right]\nonumber \\
 & = & \int\text{d}t'\delta(t-t')\left[\delta(t'-t'')e^{\lambda_{b}(t'-t'')}\right.\nonumber \\
 &  & \left.+\theta(t'-t'')\lambda_{b}e^{\lambda_{b}(t'-t'')}-\lambda_{b}\theta(t'-t'')e^{\lambda_{b}(t'-t'')}\right]\nonumber \\
 & = & \delta(t-t'')
\end{eqnarray}
The only changed interaction vertex is 
\begin{eqnarray}
\diaA & = & -2\,\mu_{b}\delta(t_{1}-t_{2})\,\delta(t_{2}-t_{3})\,\left[\partial_{t_{1}}+\partial_{t_{2}}+\partial_{t_{3}}-3\lambda_{b}\right]\nonumber \\
 & \widehat{=} & 3!\,\mu_{b}\delta(t_{1}-t_{2})\,\delta(t_{2}-t_{3})\,\lambda_{b},
\end{eqnarray}
where we used in the last step that the derivatives lead to vanishing
contribution to all diagrams up to linear order in $\mu_{b}$ as we
showed in Eq.\ \ref{eq:vansishing}. The relevant diagrams correcting
the posterior mean are then 
\begin{eqnarray}
 &  & \diaD+\diaE+\diaF\nonumber \\
 & = & -3!\mu_{b}\int_{0}^{t_{\text{o}}}\text{d}t'\,D_{tt'}^{\text{b}}\left[\frac{\lambda_{b}}{2}m_{b}^{2}(t')+\frac{\lambda_{b}}{2}D_{t't'}^{\text{b}}-\underbrace{D_{t't'}^{\text{f}}}_{\nicefrac{1}{2}}\right]\nonumber \\
 & = & -3\mu_{b}\int_{0}^{t_{\text{o}}}\text{d}t'\,\left[f_{b}(t,t')-\frac{f_{b}(t,t_{\text{o}})f_{b}(t',t_{\text{o}})}{f_{b}(t_{\text{o}},t_{\text{o}})}\right]\times\nonumber \\
 &  & \left[\left[\frac{f_{b}^{2}(t',t_{\text{o}})}{f_{b}^{2}(t_{\text{o}},t_{\text{o}})}\,d^{2}+f_{b}(t',t')-\frac{f_{b}^{2}(t^{'},t_{\text{o}})}{f_{b}(t_{\text{o}},t_{\text{o}})}\right]\lambda_{b}-1\right]\nonumber \\
\end{eqnarray}
 This integral can be calculated analytically. However, the resulting
expression is relatively complicated, therefore omitted here, and
only plotted in Fig.\ \ref{fig:Uncertainty-of-a-1}. We calculate
it with the computer algebra system \texttt{SymPy} \citep{10.7717/peerj-cs.103}.
The same is true for the first order (in $\mu_{b}$) correction to
the uncertainty
\begin{eqnarray}
 &  & \diaG\nonumber \\
 & = & -3!\mu_{b}\lambda_{b}\int_{0}^{t_{\text{o}}}\text{d}t''\,D_{tt''}^{\text{b}}m_{b}(t'')D_{t''t'}^{\text{b}},\nonumber \\
\end{eqnarray}
which we also only present graphically in Fig.\ \ref{fig:Uncertainty-of-a-1}.

This figure shows that in all displayed cases ($\lambda_{b}\in\{-1,0,1\}$)
the posterior trajectories preferentially avoid getting close to easily
diverging regimes (larger positive values for $\mu_{b}>0$), and they
avoid such areas the more, as more the linear dynamics is unstable
($i.e.$ larger values of $\lambda_{b}$).

Interestingly, the interplay of this non-linear dynamics with the
constraint provided by the measurement leads to a reduced a posteriori
uncertainty for unstable systems ($\lambda_{b}>0$) for times prior
to the measurement. This is not in contradiction to the notion of
chaotic systems being harder to predict. Here, we are looking at trajectories
that could have let -- starting from some known value -- to the
observed situation at a later time. Thanks to the stronger divergence
of trajectories of chaotic systems, the variety of trajectories that
pass through both, the initial condition and the later observed situation,
is smaller than if the system is not chaotic. Thus, the measurement
provides more information for this period in the chaotic regime, but
less for the period after the measurement.

\section{Conclusion and Outlook}

\label{sec:Conclusion-and-Outlook}

We brought dynamical field inference based on information field theory
and the suspersymmetric theory of stochastics into contact. To this
end, we showed that the DFI partition function becomes the STS one
if the excitation of the field becomes white Gaussian noise and no
measurements constrain the field evolution. In this case, the dynamical
system has a supersymmetry. We note that neither STS nor DFI are limited
to the white noise case.

For chaotic systems, this supersymmetry is broken spontaneously. As
the presence of chaos limits the ability to predict a system, DFI
for systems with broken supersymmetry should become more difficult.
We hope that the here established connection of STS and DFI allows
to quantitatively investigate this.

While re-deriving basic elements of STS within the framework of IFT,
we carefully investigate the domains on which the different fields
and operators live and act, respectively, using the perspective that
the continuous time description of the system should be the limiting
case of a discrete time representation for vanishing time steps. Thereby,
we show, for example, that the fermionic ghost field has to vanish
on the initial time slice for the theory to be consistent.

Furthermore, we show that most measurements of the field during its
evolution phase do not obey the system's supersymmetry, they are not
$Q$-exact. Nevertheless, the formalism of STS is still applicable
and might help to develop advanced DFI schemes. For example, two of
the challenges DFI is facing are the representation of the dynamics
enforcing delta function and a Jacobian in the path integral of the
DFI partition function. For these STS introduces bosonic Lagrange
and fermionic ghost fields. Using those in perturbative calculations,
for example via Feynman diagrams, might allow to develop DFI schemes
that are able to cope with non-linear dynamical systems.

In order to illustrate how such a non-linear dynamics inference would
look like, we investigate a simplified situation, in which the deviation
of a system driven by stochastic external excitation from the classical
(not perturbed system) is measured at an initial and a later time.
The simplifications we impose are that (i) the measurement probes
exactly one eigenmode of the linear part of the evolution operator
for these deviations, that (ii) the evolution operator stays stationary
during the considered period (thus different modes do not mix), and
that the non-linear part of the evolution is also (iii) stationary,
(iv) second order in the observed eigenmode, and (v) keeps that mode
also separate from the other modes (no non-linear mode mixing). Under
these particular conditions (i)-(v) the field inference problem becomes
a one dimensional problem for the measured mode as a function of time,
which can be treated exactly for a vanishing non-linearity and perturbatively
with the help of Feynman diagrams in case of non-vanishing non-linearity.
Thereby, it turns out that the Fermionic contributions, which implement
the effect of the functional determinant, are key to obtain the correct
a posteriori mean of the system.

The investigation of the illustrative example show a few things. First,
predicting the future evolution of a more chaotic system from measurements
is harder than for a less chaotic one as the absolute uncertainty
of the measured mode increases faster in the former situation. This
is not very surprising, but the following insight might be: Second,
the relative uncertainty (uncertainty standard deviation over absolute
value of the deviation) grows slower for a chaotic system. This is
an echo of the known memory effect of chaotic systems, which remember
small perturbations in unstable modes for a longer time thanks to
their rapid amplification. Third, non-linear dynamics, which can lead
to even more drastic divergence of system trajectories (even to infinity
in finite times), makes prediction of the future even harder, but
enhances the amount of information measurements provide for periods
between them. Due to the larger sensitivity of the system to perturbations,
the measurements now exclude more trajectories that were possible
a priori.

Thus, the interplay of measurements and non-linear chaotic systems
is complex and more interesting phenomena should become visible as
soon as the simplifying assumptions (i)-(v) made in our illustrative
example are dropped. For those, the inclusion of the Fermionic part
of the information field theory of stochastic systems will be as essential
to obtain the correct statistics on the system trajectories as it
is in our idealized illustrative example. We believe that insights
provided by the stochastic theory of supersymmetry will continue to
pay off in investigations of more complex systems, which we leave
for future research.
\begin{acknowledgments}
We acknowledge insightful discussions with Reimar Leike, and Jens
Jasche. This work was supported partly by the Excellence Cluster Universe.
\end{acknowledgments}

\bibliography{mybib}
 \bibliographystyle{unsrt}

\appendix
\newpage{}

\section{Feynman diagrams\label{sec:Feynman-diagrams}}

Here we calculate explicitly the Feynman diagrams from Sec.\ \ref{subsec:Idealized-Non-Linear-Dynamics}
for the case $\lambda_{b}=0$. These are

\begin{eqnarray}
 &  & \diaD\nonumber \\
 & = & \frac{1}{2}2\,\mu_{b}\int_{0}^{t_{\text{o}}}\text{d}t'\,\left\{ 2D_{tt'}^{\text{b}}m_{t'}\left[\partial_{t''}\,m_{t''}\right]_{t''=t'}\right.\nonumber \\
 &  & \left.+m_{t'}^{2}\left[\partial_{t''}\,D_{tt''}^{\text{b}}\right]_{t''=t'}\right\} \nonumber \\
 & = & \mu_{b}\int_{0}^{t_{\text{o}}}\text{d}t'\,\left\{ 2D_{tt'}^{\text{b}}m_{t'}\left[\partial_{t''}\,\frac{t''d}{t_{\text{o}}}\right]_{t''=t'}\right.\nonumber \\
 &  & \left.+m_{t'}^{2}\left[\partial_{t''}\,\left(\min(t,t'')-\frac{t\,t''}{t_{\text{o}}}\right)\right]_{t''=t'}\right\} \nonumber \\
 & = & \mu_{b}\int_{0}^{t_{\text{o}}}\text{d}t'\,\left\{ 2D_{tt'}^{\text{b}}\frac{t'd^{2}}{t_{\text{o}}^{2}}+\frac{t'^{2}d^{2}}{t_{\text{o}}^{2}}\left[\theta(t-t')-\frac{t}{t_{\text{o}}}\right]\right\} \nonumber \\
 & = & \mu_{b}d^{2}\int_{0}^{t_{\text{o}}}\text{d}t'\,\left\{ 2\,\left(\min(t,t')-\frac{t\,t'}{t_{\text{o}}}\right)\frac{t'}{t_{\text{o}}^{2}}\right.\nonumber \\
 &  & \left.+\frac{t'^{2}}{t_{\text{o}}^{2}}\left[\theta(t-t')-\frac{t}{t_{\text{o}}}\right]\right\} \nonumber \\
 & = & {\color{teal}2\mu_{b}d^{2}\int_{0}^{t}\text{d}t'\frac{t'^{2}}{t_{\text{o}}^{2}}}{\color{olive}+2\mu_{b}d^{2}\int_{t}^{t_{\text{o}}}\text{d}t'\frac{t't}{t_{\text{o}}^{2}}}\nonumber \\
 &  & {\color{orange}{\color{brown}-2\mu_{b}d^{2}\int_{0}^{t_{\text{o}}}\text{d}t'\,\frac{t\,t'^{2}}{t_{\text{o}}^{3}}}}\nonumber \\
 &  & {\color{teal}+\mu_{b}d^{2}\int_{0}^{t}\text{d}t'\frac{t'^{2}}{t_{\text{o}}^{2}}}{\color{brown}-\mu_{b}d^{2}\int_{0}^{t_{\text{o}}}\text{d}t'\,\frac{t\,t'^{2}}{t_{\text{o}}^{3}}}\nonumber \\
 & = & {\color{teal}\mu_{b}d^{2}\frac{t^{3}}{t_{\text{o}}^{2}}}{\color{brown}-\mu_{b}d^{2}\,t}{\color{olive}+\mu_{b}d^{2}\frac{(t_{\text{o}}^{2}-t^{2})\,t}{t_{\text{o}}^{2}}}\nonumber \\
 & = & \mu_{b}d^{2}\frac{{\color{teal}t^{3}}{\color{brown}-t_{\text{o}}^{2}t}{\color{olive}+t_{\text{o}}^{2}t-t^{3}}}{t_{\text{o}}^{2}}=0,
\end{eqnarray}
\begin{eqnarray}
 &  & \diaE\nonumber \\
 & = & \frac{1}{2}2\,\mu_{b}\int_{0}^{t_{\text{o}}}\text{d}t'\,\left\{ 2D_{tt'}^{\text{b}}\left[\partial_{t''}\,D_{t't''}^{\text{b}}\right]_{t''=t'}\right.\nonumber \\
 &  & \left.+D_{t't'}^{\text{b}}\left[\partial_{t''}\,D_{tt''}^{\text{b}}\right]_{t''=t'}\right\} \nonumber \\
 & = & 2\mu_{b}\int_{0}^{t_{\text{o}}}\text{d}t'\,D_{tt'}^{\text{b}}\left[\partial_{t''}\,\left(\min(t',t'')-\frac{t'\,t''}{t_{\text{o}}}\right)\right]_{t''=t'}\nonumber \\
 &  & +\mu_{b}\int_{0}^{t_{\text{o}}}\text{d}t'\,D_{t't'}^{\text{b}}\left[\partial_{t''}\,\left(\min(t,t'')-\frac{t\,t''}{t_{\text{o}}}\right)\right]_{t''=t'}\nonumber \\
 & = & 2\mu_{b}\int_{0}^{t_{\text{o}}}\text{d}t'\,\left(\min(t,t')-\frac{t\,t'}{t_{\text{o}}}\right)\,\left[\underbrace{\theta(t'-t')}_{\nicefrac{1}{2}}-\frac{t'}{t_{\text{o}}}\right]\nonumber \\
 &  & +\mu_{b}\int_{0}^{t_{\text{o}}}\text{d}t'\,\left(\min(t',t')-\frac{t'^{2}}{t_{\text{o}}}\right)\,\left[\theta(t-t')-\frac{t}{t_{\text{o}}}\right]\nonumber \\
 & = & 2\mu_{b}\int_{0}^{t}\text{d}t'\,\left[\frac{t'}{2}-\frac{t'^{2}}{t_{\text{o}}}\right]+2\mu_{b}\,t\int_{t}^{t_{\text{o}}}\text{d}t'\,\left[\frac{1}{2}-\frac{t'}{t_{\text{o}}}\right]\nonumber \\
 &  & +2\mu_{b}\int_{0}^{t_{\text{o}}}\text{d}t'\,\left(-\frac{t\,t'}{2t_{\text{o}}}+\frac{t\,t'^{2}}{t_{\text{o}}^{2}}\right)\nonumber \\
 &  & +\mu_{b}\int_{0}^{t}\text{d}t'\,\left(t'-\frac{t'^{2}}{t_{\text{o}}}\right)-\mu_{b}\int_{0}^{t_{\text{o}}}\text{d}t'\,\left(t'-\frac{t'^{2}}{t_{\text{o}}}\right)\,\frac{t}{t_{\text{o}}}\nonumber \\
 & = & 2\mu_{b}\,\left(\frac{t^{2}}{4}-\frac{t^{3}}{3t_{\text{o}}}\right)+2\mu_{b}\,t\,\left[\frac{(t_{\text{o}}-t)}{2}-\frac{(t_{\text{o}}^{2}-t^{2})}{2t_{\text{o}}}\right]\nonumber \\
 &  & +2\mu_{b}\,\left(-\frac{t\,t_{\text{o}}^{2}}{4t_{\text{o}}}+\frac{t\,t_{\text{o}}^{3}}{3t_{\text{o}}^{2}}\right)\nonumber \\
 &  & +\mu_{b}\,\left(\frac{t^{2}}{2}-\frac{t^{3}}{3t_{\text{o}}}\right)-\mu_{b}\,\left(\frac{t_{\text{o}}^{2}}{2}-\frac{t_{\text{o}}^{3}}{3t_{\text{o}}}\right)\,\frac{t}{t_{\text{o}}}\nonumber \\
 & = & \frac{\mu_{b}}{6}\,\left({\color{brown}3t^{2}}{\color{purple}-\frac{4t^{3}}{t_{\text{o}}}}+6t\,({\color{teal}t_{\text{o}}}{\color{brown}-t})-\frac{6t\,({\color{teal}t_{\text{o}}^{2}}{\color{purple}-t^{2}})}{t_{\text{o}}}\right.\nonumber \\
 &  & \left.{\color{teal}-3t\,t_{\text{o}}+4t\,t_{\text{o}}}{\color{brown}+3t^{2}}{\color{purple}-\frac{2t^{3}}{t_{\text{o}}}}{\color{teal}-3t\,t_{\text{o}}+2tt_{\text{o}}}\right)\nonumber \\
 & = & \frac{\mu_{b}}{6}\,\left({\color{brown}0}{\color{purple}+0}{\color{teal}+0}\right)=0,
\end{eqnarray}
using here and in the following color coding to highlight canceling
terms,
\begin{eqnarray}
 &  & \diaF\nonumber \\
 & = & -3!\,\mu_{b}\int_{0}^{t_{\text{o}}}\text{d}t'\,D_{tt'}^{\text{b}}D_{t't'}^{\text{f}}\nonumber \\
 & = & -6\,\mu_{b}\int_{0}^{t_{\text{o}}}\text{d}t'\,\left(\min(t,t')-\frac{t\,t'}{t_{\text{o}}}\right)\,\underbrace{\theta(t'-t')}_{\nicefrac{1}{2}}\nonumber \\
 & = & -3\,\mu_{b}\left[\int_{0}^{t}\text{d}t'\,t'+\int_{t}^{t_{\text{o}}}\text{d}t'\,t-\int_{0}^{t_{\text{o}}}\text{d}t'\,\frac{t\,t'}{t_{\text{o}}}\right]\nonumber \\
 & = & -3\,\mu_{b}\left[\frac{t^{2}}{2}+(t_{\text{o}}-t)\,t-\,\frac{t\,t_{\text{o}}}{2}\right]\nonumber \\
 & = & -\frac{3}{2}\,\mu_{b}t\,(t_{\text{o}}-t),
\end{eqnarray}
and (assuming $0\le t\le t'\le t_{\text{o}}$)

\begin{eqnarray}
 &  & \diaG\nonumber \\
 & = & 2\,\mu_{b}\int_{0}^{t_{\text{o}}}\text{d}t''\,\left\{ \left[\left[\partial_{t''}D_{tt''}^{\text{b}}\right]m_{b}(t'')D_{t''t'}^{\text{b}}\right]\right.\nonumber \\
 &  & \left.+D_{tt''}^{\text{b}}\left[\partial_{t''}m_{b}(t'')\right]D_{t''t'}^{\text{b}}+D_{tt''}^{\text{b}}m_{b}(t'')\left[\partial_{t''}D_{t''t'}^{\text{b}}\right]\right\} \nonumber \\
 & = & 2\,\mu_{b}\int_{0}^{t_{\text{o}}}\text{d}t''\,\left\{ \left[\left(\theta(t-t'')-\frac{t}{t_{\text{o}}}\right)\frac{t''d}{t_{\text{o}}}\,\left(\min(t'',t')-\frac{t''\,t'}{t_{\text{o}}}\right)\right]\right.\nonumber \\
 &  & +\left(\min(t,t'')-\frac{t\,t''}{t_{\text{o}}}\right)\frac{d}{t_{\text{o}}}\left(\min(t'',t')-\frac{t''\,t'}{t_{\text{o}}}\right)\nonumber \\
 &  & \left.+\left(\min(t,t'')-\frac{t\,t''}{t_{\text{o}}}\right)\frac{t''d}{t_{\text{o}}}\left(\theta(t'-t'')-\frac{t'}{t_{\text{o}}}\right)\right\} \nonumber \\
 & = & 2\mu_{b}\biggl\{\int_{0}^{t}\text{d}t''\,\biggl[\biggl(1-\frac{t}{t_{\text{o}}}\biggr)\frac{t''d}{t_{\text{o}}}\biggl(t''-\frac{t''t'}{t_{\text{o}}}\biggr)\nonumber \\
 &  & +\biggl(t''-\frac{t''t}{t_{\text{o}}}\biggr)\frac{d}{t_{\text{o}}}\biggl(t''-\frac{t''t'}{t_{\text{o}}}\biggr)+\biggl(t''-\frac{t''t}{t_{\text{o}}}\biggr)\frac{t''d}{t_{\text{o}}}\biggl(1-\frac{t'}{t_{\text{o}}}\biggr)\biggr]\nonumber \\
 &  & +\int_{t}^{t'}\text{d}t''\,\biggl[\biggl(-\frac{t}{t_{\text{o}}}\biggr)\frac{t''d}{t_{\text{o}}}\biggl(t''-\frac{t''t'}{t_{\text{o}}}\biggr)\nonumber \\
 &  & +\biggl(t-\frac{t''t}{t_{\text{o}}}\biggr)\frac{d}{t_{\text{o}}}\biggl(t''-\frac{t''t'}{t_{\text{o}}}\biggr)+\biggl(t-\frac{t''t}{t_{\text{o}}}\biggr)\frac{t''d}{t_{\text{o}}}\biggl(1-\frac{t'}{t_{\text{o}}}\biggr)\biggr]\nonumber \\
 &  & +\int_{t'}^{t_{\text{o}}}\text{d}t''\,\biggl[\biggl(-\frac{t}{t_{\text{o}}}\biggr)\frac{t''d}{t_{\text{o}}}\biggl(t'-\frac{t''t'}{t_{\text{o}}}\biggr)\nonumber \\
 &  & +\biggl(t-\frac{t''t}{t_{\text{o}}}\biggr)\frac{d}{t_{\text{o}}}\biggl(t'-\frac{t''t'}{t_{\text{o}}}\biggr)+\biggl(t-\frac{t''t}{t_{\text{o}}}\biggr)\frac{t''d}{t_{\text{o}}}\biggl(-\frac{t'}{t_{\text{o}}}\biggr)\biggr]\biggr\}\nonumber \\
 & = & 2\mu_{b}\biggl\{\int_{0}^{t}\text{d}t''\,\biggl[\frac{3t''^{2}d}{t_{\text{o}}}-\frac{3t''^{2}dt'}{t_{\text{o}}^{2}}-\frac{3t''^{2}dt}{t_{\text{o}}^{2}}+\frac{3t''^{2}dtt'}{t_{\text{o}}^{3}}\biggr]\nonumber \\
 &  & +\int_{t}^{t'}\text{d}t''\,\biggr[-\frac{3t''^{2}dt}{t_{\text{o}}^{2}}+\frac{3t''^{2}dtt'}{t_{\text{o}}^{3}}+\frac{2t''dt}{t_{\text{o}}}-\frac{2t''^{2}dtt'}{t_{\text{o}}^{2}}\biggr]\nonumber \\
 &  & \left.+\int_{t'}^{t_{\text{o}}}\text{d}t''\,\biggl[-\frac{4t''\text{d}tt'}{t_{\text{o}}^{2}}+3\frac{t''^{2}\text{d}tt'}{t_{\text{o}}^{3}}+\frac{\text{d}tt'}{t_{\text{o}}}\biggr]\right\} \nonumber \\
 & = & 2\mu_{b}\biggl\{\biggr[{\color{violet}\frac{t^{3}d}{t_{\text{o}}}}{\color{green}-\frac{t^{3}dt'}{t_{\text{o}}^{2}}}{\color{gray}-\frac{t^{4}d}{t_{\text{o}}^{2}}}{\color{olive}+\frac{t^{4}dt'}{t_{\text{o}}^{3}}\biggr]}+\biggl[{\color{orange}-\frac{t'^{3}dt}{t_{\text{o}}^{2}}}{\color{gray}+\frac{t^{4}d}{t_{\text{o}}^{2}}}\nonumber \\
 &  & {\color{blue}+\frac{t'^{4}dt}{t_{\text{o}}^{3}}}{\color{olive}-\frac{t^{4}dt'}{t_{\text{o}}^{3}}}{\color{cyan}+\frac{t'^{2}dt}{t_{\text{o}}}}{\color{violet}-\frac{t^{3}d}{t_{\text{o}}}}{\color{orange}-\frac{t'^{3}dt}{t_{\text{o}}^{2}}}{\color{green}+\frac{t^{3}dt'}{t_{\text{o}}}}\biggr]\nonumber \\
 &  & \left.+\biggl[{\color{purple}-2dtt'}{\color{orange}+2\frac{t'^{3}dt}{t_{\text{o}}^{2}}}{\color{purple}+dtt'}{\color{blue}-\frac{t'^{4}dt}{t_{\text{o}}^{3}}}{\color{purple}+dtt'}{\color{cyan}-\frac{t'^{2}dt}{t_{\text{o}}}}\biggr]\right\} \nonumber \\
 & = & 0.
\end{eqnarray}

\end{document}